\documentclass[journal,twoside]{IEEEtran}
\usepackage{amsmath,amsfonts,amssymb}
\usepackage{algorithmic}
\usepackage{algorithm}
\usepackage{array}
\usepackage[caption=false,font=normalsize,labelfont=sf,textfont=sf]{subfig}
\usepackage{textcomp}
\usepackage{stfloats}
\usepackage{url}
\usepackage{verbatim}
\usepackage{graphicx}
\usepackage{cite}
\usepackage{enumitem}
\usepackage{tikz}

\def\BibTeX{{\rm B\kern-.05em{\sc i\kern-.025em b}\kern-.08em
		T\kern-.1667em\lower.7ex\hbox{E}\kern-.125emX}}
\usepackage{balance}

\newenvironment{proof}{\noindent{\em \textbf{Proof.}}}{\quad \hfill$\Box$\vspace{2ex}}

\newtheorem{theorem}{Theorem}[section]
\newtheorem{definition}[theorem]{Definition}

\newtheorem{example}[theorem]{Example}

\newtheorem{Method}[theorem]{Method}

\def\qi {\mathbf{i}}
\def\qj {\mathbf{j}}
\def\qk {\mathbf{k}}
\def\i {\mathbf{i}}
\def\j {\mathbf{j}}
\def\k {\mathbf{k}}
\def\e {\mathbf e}
\def\ux{\underline{x}}

\def\uy{\underline{y}}

\def\R{\mathbb{R}}

\ifCLASSINFOpdf
\else
\fi

\begin{document}
	\title{The Color Clifford Hardy Signal: Application to Color Edge Detection and Optical Flow}
	\author{Xiaoxiao~Hu,\,\, Kit~Ian~Kou,\,\,Cuiming~Zou,\,\,Dong~Cheng 	
		\thanks{Xiaoxiao~Hu is with the First Affiliated Hospital of Wenzhou Medical University, Wenzhou Medical University, Wenzhou 325035, China (e-mail: huxiaoxiao@wmu.edu.cn).}
		\thanks{Kit~Ian~Kou is with the Department of Mathematics, Faculty of Science and Technology, University of Macau, Macau 999078, China  (e-mail: kikou@umac.mo).}
		\thanks{Cuiming~Zou is with the College of Science, Huazhong Agricultural University, Wuhan 430070, China  (e-mail: zoucuiming2006@163.com).}
		\thanks{Dong~Cheng is with the Department of Mathematics, Faculty of Arts and Sciences, Beijing Normal University, Zhuhai 519087, China (e-mail: chengdong720@163.com).}
	}
	
	
	\markboth{Journal of \LaTeX\ Class Files,~Vol.~14, No.~8, August~2021}%
	{Hu \MakeLowercase{\textit{et al.}}: The Color Clifford Hardy Signal: Application to Color Edge Detection and Optical Flow}
	
	
	\maketitle
	
	\begin{abstract}
		This paper introduces the idea of the color Clifford Hardy signal, which can be used to process color images. As a complex analytic function's high-dimensional analogue, the color Clifford Hardy signal  inherits many desirable qualities of analyticity. A crucial tool for getting the color and structural data is the local feature representation of a color image in the color Clifford Hardy signal. By looking at the extended Cauchy-Riemann equations in the high-dimensional space, it is possible to see the connection between the different parts of the color Clifford Hardy signal. Based on the distinctive and important local amplitude and local phase generated by the color Clifford Hardy signal, we propose five methods to identify the edges of color images with relation to a certain color.    To prove the superiority of the offered methodologies, numerous comparative studies employing image quality assessment criteria are used. Specifically by using the multi-scale structure of the color Clifford Hardy signal, the proposed approaches are resistant to a variety of noises. In addition, a color optical flow  detection method with anti-noise ability is provided as an example of  application.
	\end{abstract}
	
	\begin{IEEEkeywords}
		Clifford algebras,  Quaternion,  Analytic signal, Edge detection,  Poisson operator.
	\end{IEEEkeywords}
	
	\section{Introduction}
	
	\IEEEPARstart{T}{he} goal of image edge detection is to locate edges where the brightness of the image abruptly changes. It is a useful technique for image processing and computer vision, especially for object segmentation and feature extraction \cite{xu2022weakly}.  
	The K-means clustering approach \cite{arthur2007k}, nearest neighbor method \cite{laaksonen1996classification}, principal component analysis of quaternions \cite{shi2007quaternion}, exchange market algorithm \cite{sathya2021color}, Gaussian process regression \cite{burke2021edge} are typical   approaches for segmenting and detecting image edges. At present, the most popular frameworks employed to detect image edges   are    convolutional neural network models \cite{feng2021deep,meng2023multiscale,chen2023edge}. These approaches do not directly model the edge structure of an image, but instead rely on huge quantities of high-quality, annotated training data to be successful with a particular kind of image problem.
	Therefore, working with edge-based segmentation models, which can segment an image without training data and maybe offer a functional form of a certain edge inside an image, is still beneficial for many exciting applications.

	In this study, we provide multiple approaches  based on the color Clifford Hardy signal (CCHS) for detecting edges of color images. The CCHS can extract color edge features in a comprehensive way since it is a higher dimensional generalization of the complex analytic function from the two-dimensional (2D) space to the six-dimensional (6D) space.


	B{\"u}low et al. \cite{bulow1999novel} provided the first definition of the quaternion analytic signal in 1999. By considering the partial Hilbert transform and total Hilbert transform for a real-valued signal, respectively, Ernstein et al. \cite{bernstein2013generalized} defined 2D quaternion analytic signal coupled with two-sided quaternion Fourier transform. The quaternion analytic signal can be described in polar form under certain assumptions, and then representations of signal properties, such as local amplitude, local phase, and local orientation, are offered.
	The energy information of a signal is revealed by the local amplitude, the structural information is provided by the local phase, and the geometric information is shown by the local orientation.
	Additionally, they demonstrated how the total Hilbert transform may serve as a precise corner detector and how partial Hilbert transformations can identify an image's edges. In the literature \cite{pei2008short}, the 2D Hilbert transform's application to edge detection is also demonstrated.
	Kou et al. \cite{kou2017envelope} constructed the 2D generalized quaternion analytic signal associated with quaternion linear canonical transform (QLCT) using the QLCT based 2D quaternion Hilbert transform. The quaternion analytic signal is useful to detect the features of images. For example, it is possible to identify the envelope of a color image by measuring the local amplitude of the 2D generalized quaternion analytic signal. The 2D quaternion Hilbert transform combined with the quaternion linear canonical transform was utilized by \cite{kou2020plancherel} to identify the edges of color images.

	As a  2D generalization of the analytic signal, the concept of monogenic signal was first developed by Felsberg and Sommer in 2001 \cite{felsberg2001monogenic}. The Riesz transform, a key idea in Clifford analysis, serves as the foundation for the definition of monogenic signal. They proposed the monogenic scale-space in   \cite{felsberg2004monogenic}, which provides the local phase-vector and attenuation in the scale-space. As a further method for edge detection, they also proposed the notion of differential phase congruency. Yang et al. provided the generalized Cauchy-Riemann equation in \cite{yang2018edge} for the functions of the monogenic scale-space, and they also suggested the modified differential phase congruency approach to improve the effectiveness of edge detection. Because the latter corresponds to the former with zero scale, every function of the monogenic scale-space is an extension of a monogenic signal. The improved differential phase congruency edge detection approach has been demonstrated in \cite{yang2018edge} to have a strong noise immunity at several  scales.


	The quaternion analytic signal was upgraded to include the quaternion Hardy function with two scales in \cite{hu2018phase}. Two different forms of phase-based edge detectors were proposed using the quaternion Hardy function's phase as a determining factor. The edge detectors based on the quaternion Hardy function give higher performance in terms of noise immunity than the techniques in \cite{felsberg2004monogenic} and \cite{yang2018edge}.  	However, the edge detectors   mentioned above \cite{kou2017envelope,kou2020plancherel,felsberg2004monogenic,yang2018edge,hu2018phase}   are only useful for gray-scale images..

Inspired by the idea of color monogenic signal \cite{demarcq2011color}, we  propose CCHS to process color images directly. 
In the current study, three 4D quaternion Hardy functions are embedded into the 6D Clifford space.
	Then, using two partial and one total Hilbert transforms, we are able to obtain the CCHS. It has more structure information than the previously stated hyper-complex function models for color images since it includes both the structure and color information of a color image. The CCHS has a strong noise immunity capabilities because of the two scales that correlate to the horizontal direction and the vertical direction, respectively. Additionally, through studies, we show that the CCHS is resistant to noises other than Gaussian noise.
	
	\subsection{Paper Contributions}
	The following is a summary of this paper's contributions.
	\begin{itemize}
		\item  Three quaternion Hardy functions are inserted into the Clifford space to define the CCHS. The CCHS may be used to directly process color images because it contains both the structure and color information of a color image. Two scale factors in the CCHS are crucial in the denoising process.
		\item The relationships between the CCHS's components   are discovered. They are important in color image processing both theoretically and practically.
		\item There are five alternative edge detection techniques suggested. The comparative studies demonstrate how resistant the suggested approaches are to various noises, particularly to Poisson and Gaussian noises.
	\end{itemize}

	\subsection{Paper Outlines}
	The remainder of this work is structured as follows. Quaternion algebra, quaternion analytical signal,   quaternion Hardy function and some of their   fundamental properties are covered in Section \ref{hxx2}.
	The major findings are developed in Section \ref{hxx3}. In particular, the CCHS is built. Five edge detection techniques are suggested based on the CCHS's attributes.
	In Section \ref{hxx4}, extensive experiments for the suggested  color edge detection approaches   are presented. Section \ref{hxx5} concludes this work by providing a summary.
	

	\section{Preliminaries}\label{hxx2}
	\subsection{Clifford and Quaternion Algebra}
	Everything that follows will be done in the real {\it Clifford algebra}, denoted by ${Cl}_{m,0}$. The majority of the fundamental concepts and notations related to Clifford algebra can be found in \cite{brackx1982clifford}. 
	
	Let \[\ {\mathbb{R}^m} =\{\underline{x} \; |\; \underline{x}=x_1 {\e}_1 + \cdots + x_m {\e}_m, x_j
	\in \R,  1\leq j\leq m \}.\] An element ${\mathbb{R}}^m$ in is referred to as a {\it vector} and must be identical to the standard Euclidean space. Additionally, let $$
	\mathbb{R}_1^m =\{x \; |\; x=x_0+\underline{x}, x_0 \in \R, \ux \in {\mathbb{R}}^m \}$$ be the {\it para-vector} space and an element in ${\mathbb{R}}_1^m $ is referred to as a {\it para-vector}.
	The multiplication of two para-vectors
	$x_0+\ux=\sum_{j=0}^{m}x_j {\e}_j$ and $y_0+\uy=\sum_{j=0}^{m}y_j {\e}_j$
	is given by
	$(x_0+\ux)(y_0+\uy)=(x_0y_0+\ux\cdot \uy) +(x_0 \uy+y_0\ux)+(\ux\wedge\uy)$
	with
	$
	\ux\cdot \uy=\langle \ux, \uy\rangle=\sum_{j=1}^{m}x_jy_j
	$
	and
	$\ux\wedge\uy=\sum_{i<j}{\e}_{ij}(x_iy_j-x_jy_i).$
	The equation $(x_0+\ux)(y_0+\uy)$ has three parts. They are designated as follows:
	\begin{itemize}
		\item The {\it scalar part}: $x_0y_0+\ux\cdot \uy={\rm{Sc}}[(x_0+\ux)(y_0+\uy)]$;
		\item the {\it vector part }: $x_0 \uy+y_0\ux={\rm{Vec}}[(x_0+\ux)(y_0+\uy)]$;
		\item the {\it bi-vector part }: $\ux\wedge\uy={\rm{Bi}}[(x_0+\ux)(y_0+\uy)]$.
	\end{itemize}

	Let  $\mathbb{H}$ be  the {\it Hamiltonian skew field of quaternions}, which
	is isomorphic to ${Cl}_{2,0}$. It has been demonstrated that quaternions offer a suitable framework for a unified approach to three- and four-dimensional signal processing \cite{jiang2020controllability,xia2020penalty}.
	A quaternionic number takes a form
	\begin{equation}\label{11}
		q:=q_0+{\qi}q_1+{\qj}q_2+{\qk}q_3,
	\end{equation}
	where $q_k\in\mathbb{R}, k=0,1,2,3,$  imaginary units $\{\qi, \qj, \qk \}$ obey the Hamilton's multiplication rules: $\qi^2=  \qj^2= \qk^2= \qi\qj\qk=-1.$

	
	There are various Quaternion Fourier Transforms (QFTs)  \cite{hu2017quaternion}. Here, we employ the following two-sided QFT  (hereinafter referred to as QFT). Let $f\in L^{1}(\mathbb{R}^2, \mathbb{H})$, its QFT is defined by
  \begin{equation*} 
	\mathcal{F}[f](\omega_1, \omega_2):=\int_{\mathbb{R}^2} e^{-{ \qi}\omega_1 x}f(x, y)e^{{- \qj} \omega_2 y}dxdy.\label{hu24}
	\end{equation*}

	\subsection{Quaternion Analytic  Signal (QAS) and  Quaternion Hardy Space}
	There are several ways to generalize the notion of analyticity to  higher-dimensional spaces \cite{yang2018edge,hu2018phase}. In this research, we explore the higher-dimensional analytical functions using the methodology described in \cite{hu2018phase}. Let's go over the definitions and characteristics of the QFT-related partial and total quaternion Hilbert transforms.
	\begin{definition}\label{def1} \cite{yang2018edge}
		The quaternion  partial Hilbert transform  (QPHT) $\mathcal{H}_{1}$ of $f$ along the $x$-axis or  $y$-axis, and the quaternion  total Hilbert transform  QTHT $\mathcal{H}_{2} $ along the $x, y$ axes of $ f$ are given by
		\begin{align} 
			\mathcal{H}_{1}[f(\cdot,y)](x)&:= \frac{1}{\pi} \mathrm{p.v.}\int_{\mathbb{R}}\frac{f(t,y)}{x-t}dt, \label{deH1}\\
			\mathcal{H}_{1}[f(x,\cdot)](y)&:= \frac{1}{\pi}\mathrm{p.v.}\int_{\mathbb{R}}\frac{f(x,t)}{y-t}dt,\label{deH2}\\
			\mathcal{H}_{2}[f(\cdot,\cdot)](x,y)&:= \frac{1}{\pi^{2}}\mathrm{p.v.}\int_{\mathbb{R}^{2}}\frac{f(t,s)}{(x-t)(y-s)}dtds.\label{H3}
		\end{align}
		Here, $f$  is a general quaternionic function such that Eqs.  (\ref{deH1}),  (\ref{deH2}) and (\ref{H3})  are well defined.
	\end{definition}


	It is easy to obtain that \cite{hu2018phase}
	\begin{align*} 
		\mathcal{F}[\mathcal{H}_{1}[f(\cdot,y)](x)](\omega_{1},\omega_{2})&= \frac{-1}{|\omega_{1} |} \mathcal{ F}[\partial f/ \partial x](\omega_{1},\omega_{2}),\\
		\mathcal{F}[\mathcal{H}_{1}[f(x,\cdot)](y)](\omega_{1},\omega_{2})&= \frac{-1}{|\omega_{2} |} \mathcal{ F}[\partial f /\partial y](\omega_{1},\omega_{2}),\\
		\mathcal{F}[\mathcal{H}_{2}[f]](\omega_{1},\omega_{2})&= \frac{1}{|\omega_{1} \omega_{2}|} \mathcal{ F}[\partial ^{2}f/\partial x\partial y](\omega_{1},\omega_{2}).
	\end{align*}
	The aforementioned equations thus demonstrate that the QPHT $\mathcal{H}_{1}$ and QTHT $\mathcal{H}_{2}$ are similar to the differentiation operation that can separate the edges. Additionally, because of their lengthy impulse responses, the QPHT and QTHT are well suited for detecting ramp edges and can lessen the impact of noise on edge detection.
	
	The composite of the original signal and its ``quaternion Hilbert transform" (a mix of the quaternion partial and total Hilbert transforms) in high  dimensional spaces was employed to define the quaternion analytic signal.
	
	\begin{definition}[Quaternion Analytic Signal] \cite{hu2018phase} \label{qftAD1}
		The QAS  of $f$ is defined  by
		\begin{align*} 
			f_{q}(x,y):=&f(x,y)+\qi\mathcal{H}_{1}[f(\cdot,y)](x)+\mathcal{H}_{1}[f(x,\cdot)](y)\qj\\
			&+\i\mathcal{H}_{2}[f](x,y)\qj,
		\end{align*}
		where $f(x,y)$ is a quaternionic function such  that $ f_{q}$ is well defined (i.e.,  Eqs. $(\ref{deH1}), (\ref{deH2})$ and $ (\ref{H3})$ are well defined).
	\end{definition}
	
	The quaternion Hardy space is a generalization of the idea of Hardy space to the multidimensional space under the quaternion situation.
	
	
	\begin{definition} [Quaternion  Hardy   Space  $\mathbb{Q}(\mathbb{C}_{\qi \qj}^{+}, \mathbb{H})$] \cite{hu2018phase} \label{QHSS}  Quaternion  Hardy   Space  $\mathbb{Q}(\mathbb{C}_{\qi \qj}^{+}, \mathbb{H})$ is the class of quaternion Hardy functions (QHF)
		$g(z_1, z_2)$ defined on the upper half space
		$ \mathbb{C}_{\qi \qj}^{+}$
		which satisfies the  following conditions.
		\begin{enumerate} 
			\item $\frac{\partial}{\partial \overline{z_{1}}}g=0,$
			$ g\frac{\partial}{\partial \overline{z_{2}}}=0, $
			\item $\int_{\mathbb{R}^2} |g(x_{1}+{\qi} y_{1}, x_{2}+{\qj} y_{2})|^{2}dx_{1}dx_{2} <\infty,$ for all $ y_{1}>0, y_{2}>0 $,
		\end{enumerate}
		where  $ \mathbb{C}_{\qi \qj}^{+}:= \{( z_1, z_2)|z_1:=x_1+\qi y_1, z_2:=x_2+\qj y_2, y_1>0, y_2 > 0\}. $  Due to the non-commutativity of   quaternions, the first equation should be understood as $\frac{\partial}{\partial \overline{z_{1}}} $ being applied from the left, and the second  equation should be understood as $\frac{\partial}{\partial \overline{z_{2}}} $ being applied from the right.
		$y_1$ and $y_2$ can be seen as scales.
	\end{definition}
	
	The QAS  corresponds to the boundary value of a quaternion Hardy function in $\mathbb{C}_{\qi \qj}^{+}.$
	
	\begin{theorem}\cite{hu2018phase}  \label{qqas}
		A quaternionic signal $f_{q}\in L^{2}(\mathbb{R}^{2}, \mathbb{H}),$ is the QAS of a  quaternionic square integrable function  if and only if $f_{q}$ is the boundary value of a quaternion Hardy function in $\mathbb{C}_{\qi \j}^{+}.$
		Namely, there exits a QHF  $\widetilde{f}_{q}(z_{1},z_{2}) \in \mathbb{Q}(\mathbb{C}_{\qi \j}^{+}, \mathbb{H}),$  such that
		\begin{align*} 
			\widetilde{f}_{q}(z_{1},z_{2})=&\frac{1}{2\pi \i} \int_{\mathbb{R}^{2}} \frac{4f(s,t)}{(s-z_{1})(t-z_{2})}dsdt\frac{1}{2\pi \j}\\
			=&u+\qi v_1 + v_2 \qj +\qi v_3 \j,
		\end{align*}
		and
		\begin{equation}\label{qaf}
			f_{q}(x_{1}, x_{2})=\lim_{y_{1}\rightarrow 0^{+},y_{2}\rightarrow 0^{+}} \widetilde{f}_{q}(x_1+\qi y_1, x_2+\qj y_2).
		\end{equation}
		That is to say,
		\begin{equation*}\label{qaf1} 
			f(x_{1}, x_{2})=\lim_{y_{1}\rightarrow 0^{+},y_{2}\rightarrow 0^{+}} u(x_1+\qi y_1,x_2+\qj y_2),
		\end{equation*}
		\begin{equation*}\label{qaf2} 
			\mathcal{H}_{1}[f(\cdot,x_2)](x_1)=\lim_{y_{1}\rightarrow 0^{+},y_{2}\rightarrow 0^{+}}v_{1}(x_1+\qi y_1, x_2+\qj y_2),
		\end{equation*}
		\begin{equation*}\label{qaf3} 
			\mathcal{H}_{1}[f(x_1,\cdot)](x_2)=\lim_{y_{1}\rightarrow 0^{+},y_{2}\rightarrow 0^{+}}v_{2}(x_1+\qi y_1, x_2+\qj y_2),
		\end{equation*}
		\begin{equation*}\label{qaf4} 
			\mathcal{H}_{2}[f](x_1,x_2)=\lim_{y_{1}\rightarrow 0^{+},y_{2}\rightarrow 0^{+}}v_{3}(x_1+\qi y_1, x_2+\qj y_2),
		\end{equation*}
		where  the functions $u$ and $v_{i}, i=1,2,3$ are constructed by the Poisson $P_{y }(x ): = \frac{y }{\pi(y ^2+x ^2)}$ and the conjugate Poisson $Q_{y}(x): = \frac{x}{\pi(y^2+x^2)}$ integrals, respectively.  That is,
		\begin{align}\label{poisson} 
			u(x_1+ & \qi y_1, x_2+\qj y_2)= [f*K^{PP}_{y_1,y_2}](x_1,x_2)\nonumber \\
			&=\frac{1}{\pi^2}\int_{\R^2}\frac{y_1 y_2 f(s,t)}{(y_{1}^2+(x_{1}-s)^2)(y_{2}^2+(x_{2}-t)^2)}dsdt\nonumber \\
			v_{1}(x_1+ &\qi y_1,x_2+\qj y_2)=  [f*K^{QP}_{y_1,y_2}](x_1,x_2)\nonumber\\ &=\frac{1}{\pi^2}\int_{\R^2}\frac{x_1 y_2 f(s,t)}{(y_{1}^2+(x_{1}-s)^2)(y_{2}^2+(x_{2}-t)^2)}dsdt,\nonumber\\
			v_{2}(x_1+ &\qi y_1, x_2+\qj y_2)= [f*K^{PQ}_{y_1,y_2}](x_1,x_2)\nonumber\\ &=\frac{1}{\pi^2}\int_{\R^2}\frac{y_1 x_2 f(s,t)}{(y_{1}^2+(x_{1}-s)^2)(y_{2}^2+(x_{2}-t)^2)}dsdt,\nonumber\\
			v_{3}(x_1+&\qi y_1, x_2+\qj y_2)= [f*K^{QQ}_{y_1,y_2}] (x_1,x_2)\nonumber\\
			& =\frac{1}{\pi^2}\int_{\R^2}\frac{x_1 x_2 f(s,t)}{(y_{1}^2+(x_{1}-s)^2)(y_{2}^2+(x_{2}-t)^2)}dsdt,\nonumber\\
		\end{align}
		where
		\begin{align*} 
			& K^{PP}_{y_1,y_2}(x_1,x_2)=P_{y_1 }(x_1 )P_{y_2 }(x_2) , \\ &K^{QP}_{y_1,y_2}(x_1,x_2)=Q_{y_1 }(x_1 )P_{y_2 }(x_2),   \\
			&  K^{PQ}_{y_1,y_2}(x_1,x_2)=P_{y_1 }(x_1 )Q_{y_2 }(x_2) , \\ &K^{QQ}_{y_1,y_2}(x_1,x_2)=Q_{y_1 }(x_1 )Q_{y_2 }(x_2),
		\end{align*}
		and  $ * $ denotes the  classical  convolution  operator in  $\mathbb{R}^{2}.$
	\end{theorem}
	
	Now that the primary results are ready, we may move forward.

	\section{Main Results} \label{hxx3}
	The purpose of this work is to build a function that represents a color image and is defined on $\mathbb{R}^6$.
	The vector-valued function $f(x_1,x_2 )=\sum_{i=1}^{3}f_{i}(x_1,x_2 ){\e_i}$ can be used to model a color image.
	We first define the color Clifford analytic signal, a generalized analytic signal from 2D to 6D, using the concept of the quaternion analytic signal. The color Clifford analytic signal is expanded to the upper half-space using Poisson operators and conjugate Poisson operators, and a color Clifford Hardy function with two scales is then generated.
	
	\subsection{The Color Clifford  Hardy Signal}\label{S3}
	
	\begin{definition}[Color Clifford Analytic Signal] \label{ccas}
		Let $f(x_1,x_2 )=\sum_{i=1}^{3}f_{i}(x_1,x_2 ){\e_i} \in L^{2}\left( \mathbb{R}^2, \mathbb{R}^3 \right)$, the color Clifford analytic signal $f_{s}$ of $f$  is defined as
		\begin{align*}
			f_{s}(x_1,x_2) = & \sum_{i=1}^{3}f_{i}(x_1,x_2 ){\e_i} 
			+  \mathcal{H}_{1}[\sum_{i=1}^{3}f_i(\cdot,x_2)] {\e_4}\\
			& +   \mathcal{H}_{1}[\sum_{i=1}^{3}f_i(x_1,\cdot) ]{\e_5}
			+ \mathcal{H}_{2}[\sum_{i=1}^{3}f_i(\cdot,\cdot) ]{\e_6} .
		\end{align*}
	\end{definition}
	
	The Poisson operators and conjugate Poisson operators are then used to expand the color Clifford analytic signal to the top half-space. The color Clifford Hardy signal is defined as follows.
	
	\begin{definition}[Color Clifford Hardy Signal (CCHS)]\label{qftAD}
		The color Clifford Hardy Signal $f_{cq}$ of a color image  $f=f_1 {\e_1} +f_2{\e_2}+ f_3{\e_3}$ is defined as
		\begin{align*} 
			f_{cq}(z_{1},z_{2}):=&\sum_{i=1}^{6}A_{i}(z_{1},z_{2}){\e_{i}},
		\end{align*}
		where $z_1 =x_1+\qi y_1, z_2 =x_2+\j y_2$ and $A_i$ ($i=1,2,\cdots,6$) are defined as follows:
		\begin{align}\label{A}
			A_{i}(z_{1},z_{2}):= &[f_{i}*K^{PP}_{y_1,y_2}](x_1,x_2), \quad i=1,2,3;\nonumber \\
			A_{4}(z_{1},z_{2}):=& [(f_{1}+f_{2}+f_{3})*K^{QP}_{y_1,y_2}](x_1,x_2)
			;\nonumber \\
			A_{5}(z_{1},z_{2}):= &[(f_{1}+f_{2}+f_{3})*K^{PQ}_{y_1,y_2}](x_1,x_2)
			;\nonumber \\
			A_{6}(z_{1},z_{2}):=& [(f_{1}+f_{2}+f_{3})*K^{QQ}_{y_1,y_2}](x_1,x_2).
		\end{align}
		Here, $A_{i} $  with $i=1,2,3$  are the smoothed color  representations for $f$ and  $ A_{i}$ with $i=4,5,6$ are the smoothed vertical and horizontal structures for $f$.
	\end{definition}
	
	The relationships between the parts of $f_{cq}$ are explained by the following theorem. In other words, the color Clifford Hardy signal's generalized Cauchy Riemann equation is established.
	
	\begin{theorem} \label{crlem}
		If $f(x_1,x_2 )=\sum_{i=1}^{3}f_{i}(x_1,x_2 ){\e_i} \in L^{2}(\mathbb{R}^{2},\mathbb{R}^3)$ and its  corresponding CCHS is
		\begin{align*}  
			f_{cq}(z_{1},z_{2}):= \sum_{i=1}^{6}A_{i}(z_{1},z_{2}){\e_{i}}.
		\end{align*}
		Then the relations of the components of $f_{cq}(z_{1},z_{2})$ are as follows:
		\begin{align*}
			&\frac{\partial(A_1+A_2+A_3)}{\partial x_1}= \frac{\partial(A_4)}{\partial y_1}, \:\;
			\frac{\partial(A_1+A_2+A_3)}{\partial x_2}=\frac{\partial(A_5)}{\partial y_2},\\
			&\frac{\partial(A_4)}{\partial x_1}= -\frac{\partial(A_1+A_2+A_3)}{\partial y_1}, \quad
			\frac{\partial(A_4)}{\partial x_2}=\frac{\partial(A_6)}{\partial y_2},\\
			&\frac{\partial(A_5)}{\partial x_1}= \frac{\partial(A_6)}{\partial y_1},\qquad \quad
			\frac{\partial(A_5)}{\partial x_2}=-\frac{\partial(A_1+A_2+A_3)}{\partial y_2},\\
			&\frac{\partial(A_6)}{\partial x_1}= -\frac{\partial(A_5)}{\partial y_1},\qquad \qquad
			\frac{\partial(A_6)}{\partial x_2}=-\frac{\partial(A_4)}{\partial y_2}.
		\end{align*}
	\end{theorem}
	The proof of Theorem \ref{crlem} can be founded in  Appendix \ref{a_sec1}.
	
	Since a color model often contains three channels, like the RGB model. Therefore, a Clifford number can be used to represent any color.
	Let $ \nu=a {\e_1}+b{\e_2}+c{\e_3}, (a,b,c \in \mathbb{R})$ be a color in the color image $f$, then the product of $f_{cq}$ and $\nu$  is given by
	\begin{equation*} 
		f_{cq}\nu={\rm{Sc}}[f_{cq}\nu] +{\rm{Bi}}[f_{cq}\nu],
	\end{equation*}
	where
	\begin{align}\label{real} 
		{\rm{Sc}}[f_{cq}\nu]& =aA_1 +b A_2 +cA_3,\\
		{\rm{Bi}}[f_{cq}\nu]&=(aA_2-bA_1){\e_1}{\e_2} +(aA_3-cA_1){\e_1}{\e_3}\nonumber\\
		&+(bA_3-cA_2){\e_2}{\e_3} 	+aA_4{\e_4}{\e_1}
		+aA_5{\e_5}{\e_1}\nonumber\\
		&+aA_6{\e_6}{\e_1}
		+bA_4{\e_4}{\e_2}
		+bA_5{\e_5}{\e_2}
		+bA_6{\e_6}{\e_2}\nonumber\\
		&+cA_4{\e_4}{\e_3}+cA_5{\e_5}{\e_3}+cA_6{\e_6}{\e_3}.\nonumber
	\end{align}
	Then the polar form of   $ f_{cq}\nu$ is
	\begin{equation*} 
		f_{cq}\nu=M e^{\frac{{\rm{Bi}}[f_{cq}\nu]}{|{\rm{Bi}}[f_{cq}\nu]|} \theta},
	\end{equation*}
	where
	\begin{equation*} 
		M:= \sqrt{|{\rm{Sc}}[f_{cq}\nu]|^{2} +|{\rm{Bi}}[f_{cq}\nu]|^{2}}
	\end{equation*}
	is the local amplitude of     $f_{cq}\nu$ and
	\begin{equation}\label{phase}
		\theta:= \arctan\bigg( \frac{|{\rm{Bi}}[f_{cq}\nu]|}{|{\rm{Sc}}[f_{cq}\nu]|}\bigg)
	\end{equation}
	is the local phase of the $f_{cq}\nu$. Theta $\theta $ actually calculates the angle between the supplied color vector $\nu$ and $f_{cq}$. It calculates how similar a color-fitted pixel is to a vector that contains the selected color.
	Therefore, we suggest the following edge detection techniques employing the $ {\rm{Sc}}[f_{cq}\nu]$ and $\theta $.
	\subsection{Edge Recognition for a Specific Color}
	The color image $f$ displayed here is in the CIE $L^{*} a^{*}b^{*}$ colorimetric system. This study aims to identify the boundaries of color images of a specific color. In light of this, we present the $2$-band image function  $ \mathbf{\mathbf{g}_{f_{cq}\nu}}$ connected to the $ f_{cq}\nu $ in the form:
	\begin{equation}\label{fcq} 
		\mathbf{\mathbf{g}_{f_{cq}\nu}}(x_1,x_2,y_1,y_2) 
		\rightarrow  \begin{pmatrix}  {\rm{Sc}}[f_{cq}\nu](x_1,x_2,y_1,y_2) \\ \theta(x_1,x_2,y_1,y_2)\end{pmatrix} ,
	\end{equation}
	where $  {\rm{Sc}}[f_{cq}\nu]$ and $ \theta$ are given in Eqs. (\ref{real}) and (\ref{phase}).
	
	The gradient-based edge detection methods referenced \cite{lovelock1989tensors,cumani1991edge,evans2006morphological} rely on the metric data of the images. To offer edge detection strategies for a certain color, we combine the CCHS and the notion of the gradient technique from \cite{cumani1991edge}. The generalized gradient magnitude is the main idea behind the suggested edge detection methods. Five formulations of the generalized gradient magnitude are provided based on the CCHS (see Eqs.   (\ref{r1}), (\ref{r2}), (\ref{r3}), (\ref{r4}), (\ref{r5})). We then go into great detail about edge detection techniques.
	
	The color image where the provided color is enhanced by the CCHS construction is denoted by the symbol $\mathbf{\mathbf{g}_{f_{cq}\nu}}$.
	Since we are looking for the edges of the specified color in the image, we can calculate the differential of the color image at point $p(x_1,x_2)$ as follows:
	$$  d\mathbf{\mathbf{g}_{f_{cq}\nu}}=\sum_{i=1}^{2} \frac{\partial \mathbf{\mathbf{g}_{f_{cq}\nu}}}{\partial x_i}dx_i.$$
	The squared norm of $ d\mathbf{\mathbf{g}_{f_{cq}\nu}}$ is given by
	$$ (d\mathbf{\mathbf{g}_{f_{cq}\nu}})^2= \sum_{i=1}^{2}\sum_{k=1}^{2}\gamma_{ik}dx_idx_k,$$
	where $\gamma_{ik}= \frac{\partial  \mathbf{\mathbf{g}_{f_{cq}\nu}}}{\partial x_i}\cdot \frac{\partial \mathbf{\mathbf{g}_{f_{cq}\nu}}}{\partial x_k}$. The dot $\cdot$ indicates  the scalar product of vectors in $\mathbb{R}^{2}$.
	It can be seen that $ \left(d\mathbf{\mathbf{g}_{f_{cq}\nu}}\right)^2$  is a metric on $\mathbb{R}^{2}$ and indicates how much the image value varies in the direction of $(dx_1,dx_2)$ .  The  $ \left(d\mathbf{\mathbf{g}_{f_{cq}\nu}}\right)^2$   is also the
	first fundamental form (see  \cite{kreyszig1991differential}).
	Therefore, the   squared local contrast  of $\mathbf{\mathbf{g}_{f_{cq}\nu}}$  at point $ p(x_1,x_2)$ in  the direction of   $\mathbf{n}=(n_1,n_2)$ is defined as follows
	\begin{equation*} 
		S(p,\mathbf{n})=\sum_{i=1}^{2}\sum_{k=1}^{2}\gamma_{ik}n_in_k.
	\end{equation*}
	The maximum value $\lambda_{+}$ and the related eigenvector $ \mathbf{n}_{+}$ are then obtained by computing the eigenvalues of the $ 2 \times 2 $ matrix $[\gamma_{ik}]$ as follows
	\begin{align}		
	\lambda_{+}&=( \gamma_{11} + \gamma_{22} + \sqrt{(\gamma_{11}-\gamma_{22})^2 +4\gamma_{12}^2} )/2, \label{r1} \\
		\mathbf{n}_{+}&=  (\cos(\theta_{+}), \sin(\theta_{+})), \label{n1}\\
		\theta_{+}&=  \frac{1}{2}\arctan \frac{2\gamma_{12}}{\gamma_{11}-\gamma_{22}}+k\pi, \quad k\in\mathbb{Z}.\nonumber
	\end{align}
	The local strength of an edge \cite{tschumperle2006fast} is correlated with the eigenvalues of the matrix $[\gamma_{ik}]$.
	A local directional maximum of $\lambda_{+}$ specified in Eq. (\ref{r1}) is where an edge point $p$ is placed in terms of the description of the edge point for $m$-band images given in \cite{cumani1991edge}.
	The first edge detection technique is then obtained.
	
	\begin{Method}[Clifford Hardy Edge Detection, CHED] \label{m1}
		For a given color $\nu=a {\e_1}+b{\e_2}+c{\e_3}$, $a,b,c \in \mathbb{R}$,  by  Eq.  $(\ref{fcq})$, we obtained the $2$-band image    $\mathbf{\mathbf{g}_{f_{cq}\nu}}(x_1,x_2,y_1,y_2)$, which is  the color image enhanced by  $\nu$. The CHED method  uses the following  generalized gradient magnitude,
		\begin{equation*}
			\lambda_{+}=\left( \gamma_{11} + \gamma_{22} + \sqrt{(\gamma_{11}-\gamma_{22})^2 +4\gamma_{12}^2}\right )/2,
		\end{equation*}
		where $	\lambda_{+} $ is defined by Eq.   (\ref{r1}).
	\end{Method}
	
	For the sake of convenience, the  derivatives of $ {\rm{Sc}}[f_{cq}\nu] $ and $ \theta$ with respect to $y_1,y_2$  are termed as the scale derivatives. Analogously, the  derivatives of $ {\rm{Sc}}[f_{cq}\nu] $ and $ \theta$ with respect to $x_1,x_2$ are termed as the spatial derivatives. Theorem \ref{crlem} establishes a bridge between the scale derivatives and the spatial derivatives. It is possible to express the generalized gradient magnitude by the scale derivatives through Theorem \ref{crlem} (see Appendix \ref{a_sec2}),  then we get the second detection method as follows.
	\begin{Method}[Modified Clifford Hardy Edge Detection, MCHED] \label{m2}
		The   generalized gradient magnitude   of    MCHED   is defined by
		\begin{equation}\label{r2} 
			\widetilde{\lambda}_{+}=( \widetilde{\gamma}_{11} + \widetilde{\gamma}_{22} + \sqrt{(\widetilde{\gamma}_{11}-\widetilde{\gamma}_{22})^2 +4\widetilde{\gamma}_{12}^2} )/2.
		\end{equation}
		The definitions of $\widetilde{\gamma}_{ik}$, $1\leq i,k \leq 2$ are given in Appendix \ref{a_sec2}.
		The corresponding  gradient direction    is given by
		\begin{align*}
			\widetilde{\mathbf{n}}_{+}&= (\cos(\widetilde{\theta}_{+}), \sin(\widetilde{\theta}_{+})),\\
			\widetilde{ \theta}_{+}:&=  \frac{1}{2}\arctan \frac{2\widetilde{\gamma}_{12}}{\widetilde{\gamma}_{11}-\widetilde{\gamma}_{22}}+k\pi, \quad k\in \mathbb{Z}.
		\end{align*}
	\end{Method}

	The differential phase congruency approach for edge identification was proposed by Felsberg and Sommer in \cite{felsberg2004monogenic}.
	They demonstrated that the edge map can be produced by scaling the local phase of the monogenic signal. Yang and Kou expanded the differential phase congruency approach to higher dimensional spaces \cite{yang2018edge}. In \cite{hu2018phase}, the authors introduced an edge detection approach based on the scale derivative of the local phase of the quaternion Hardy function, which was motivated by the method of measuring the scale derivative of the local phase of the monogenic signal. Therefore, the edge points information can be found in the scale derivatives of    $\mathbf{\mathbf{g}_{f_{cq}\nu}}$.  By including the partial derivative of $\mathbf{\mathbf{g}_{f_{cq}\nu}} $ with respect to scales $y_1$ and $y_2$, respectively, the color image differential at position $p(x_1,x_2)$ is then computed as follows:
	$$  d\mathbf{\mathbf{g}_{f_{cq}\nu}}=\sum_{i=1}^{2} \frac{\partial \mathbf{\mathbf{g}_{f_{cq}\nu}}}{\partial x_i}dx_i+ \sum_{i=1}^{2} \frac{\partial \mathbf{\mathbf{g}_{f_{cq}\nu}}}{\partial y_i}dy_i.$$
	Then, the  squared norm of $ d\mathbf{\mathbf{g}_{f_{cq}\nu}}$ becomes
	\begin{equation}\label{fcq2} 
		\begin{split}
			(d\mathbf{\mathbf{g}_{f_{cq}\nu}})^2=&\sum_{i=1}^{2}\sum_{k=1}^{2}  \frac{   \partial \mathbf{\mathbf{g}_{f_{cq}\nu}} }{\partial x_i} \cdot \frac{   \partial \mathbf{\mathbf{g}_{f_{cq}\nu}} }{\partial x_k}dx_idx_k \\
			&+\sum_{i=1}^{2}\sum_{k=1}^{2}  \frac{   \partial \mathbf{\mathbf{g}_{f_{cq}\nu}} }{\partial x_i} \cdot \frac{   \partial \mathbf{\mathbf{g}_{f_{cq}\nu}} }{\partial y_k}dx_idy_k\\
			&+\sum_{i=1}^{2}\sum_{k=1}^{2}  \frac{   \partial \mathbf{\mathbf{g}_{f_{cq}\nu}} }{\partial y_i} \cdot \frac{   \partial \mathbf{\mathbf{g}_{f_{cq}\nu}} }{\partial x_k}dy_idx_k\\
			&+\sum_{i=1}^{2}\sum_{k=1}^{2}  \frac{   \partial \mathbf{\mathbf{g}_{f_{cq}\nu}} }{\partial y_i} \cdot \frac{   \partial \mathbf{\mathbf{g}_{f_{cq}\nu}} }{\partial y_k}dy_idy_k.		
		\end{split}
	\end{equation}
	
	The generalized first fundamental form can also be used to identify edges, as described in \cite{kreyszig1991differential,tschumperle2006fast}.
	The partial derivatives with respect to $ x_1, x_2, y_1, y_2$   provide the matrix representation  $I_1(p)$   of the squared norm $(d\mathbf{\mathbf{g}_{f_{cq}\nu}})^2$ at the point $p(x_1,x_2,y_1,y_2)$ in the coordinate system. The expression of $I_1(p)$   is given by (\ref{ii1}) in Page \pageref{ii1}.
	\begin{figure*}[!t] 
		\begin{equation}\label{ii1}  
			I_1(p)=
			\left(
			\begin{array}{cccc}
				\frac{\partial\mathbf{\mathbf{g}_{f_{cq}\nu}}}{\partial x_1}\cdot\frac{\partial\mathbf{\mathbf{g}_{f_{cq}\nu}}}{\partial x_1} & \frac{\partial\mathbf{\mathbf{g}_{f_{cq}\nu}}}{\partial x_1}\cdot\frac{\partial\mathbf{\mathbf{g}_{f_{cq}\nu}}}{\partial x_2} & \frac{\partial\mathbf{\mathbf{g}_{f_{cq}\nu}}}{\partial x_1}\cdot\frac{\partial\mathbf{\mathbf{g}_{f_{cq}\nu}}}{\partial y_1} & \frac{\partial\mathbf{\mathbf{g}_{f_{cq}\nu}}}{\partial x_1}\cdot\frac{\partial\mathbf{\mathbf{g}_{f_{cq}\nu}}}{\partial y_2} \\
				\frac{\partial\mathbf{\mathbf{g}_{f_{cq}\nu}}}{\partial x_2}\cdot\frac{\partial\mathbf{\mathbf{g}_{f_{cq}\nu}}}{\partial x_1} &  \frac{\partial\mathbf{\mathbf{g}_{f_{cq}\nu}}}{\partial x_2}\cdot\frac{\partial\mathbf{\mathbf{g}_{f_{cq}\nu}}}{\partial x_2} &   \frac{\partial\mathbf{\mathbf{g}_{f_{cq}\nu}}}{\partial x_2}\cdot\frac{\partial\mathbf{\mathbf{g}_{f_{cq}\nu}}}{\partial y_1} &   \frac{\partial\mathbf{\mathbf{g}_{f_{cq}\nu}}}{\partial x_2}\cdot\frac{\partial\mathbf{\mathbf{g}_{f_{cq}\nu}}}{\partial y_2} \\
				\frac{\partial\mathbf{\mathbf{g}_{f_{cq}\nu}}}{\partial y_1}\cdot\frac{\partial\mathbf{\mathbf{g}_{f_{cq}\nu}}}{\partial x_1} &    \frac{\partial\mathbf{\mathbf{g}_{f_{cq}\nu}}}{\partial y_1}\cdot\frac{\partial\mathbf{\mathbf{g}_{f_{cq}\nu}}}{\partial x_2} &    \frac{\partial\mathbf{\mathbf{g}_{f_{cq}\nu}}}{\partial y_1}\cdot\frac{\partial\mathbf{\mathbf{g}_{f_{cq}\nu}}}{\partial y_1} &    \frac{\partial\mathbf{\mathbf{g}_{f_{cq}\nu}}}{\partial y_1}\cdot\frac{\partial\mathbf{\mathbf{g}_{f_{cq}\nu}}}{\partial y_2} \\
				\frac{\partial\mathbf{\mathbf{g}_{f_{cq}\nu}}}{\partial y_2}\cdot\frac{\partial\mathbf{\mathbf{g}_{f_{cq}\nu}}}{\partial x_1} &   \frac{\partial\mathbf{\mathbf{g}_{f_{cq}\nu}}}{\partial y_2}\cdot\frac{\partial\mathbf{\mathbf{g}_{f_{cq}\nu}}}{\partial x_2} &   \frac{\partial\mathbf{\mathbf{g}_{f_{cq}\nu}}}{\partial y_2}\cdot\frac{\partial\mathbf{\mathbf{g}_{f_{cq}\nu}}}{\partial y_1} &   \frac{\partial\mathbf{\mathbf{g}_{f_{cq}\nu}}}{\partial y_2}\cdot\frac{\partial\mathbf{\mathbf{g}_{f_{cq}\nu}}}{\partial y_2}\\
			\end{array}
			\right). \tag{$\star$}
		\end{equation}
		\hrulefill
	\end{figure*}
	The following techniques, which we refer to as "Matrix and Scale Edge Detection Methods"  (MaSED), make use of the CCHS's matrix representation and scale derivatives.
	\begin{Method}[Matrix and Scale Edge Detection  Method 1, MaSED1]\label{m3}
		The following is the definition of the generalized gradient magnitude of the MaSED1:
		\begin{equation}\label{r3}
			w_1(p)=\sum_{i=1}^{4} \lambda_{1i}(p),
		\end{equation}
		where $\lambda_{11}(p), \lambda_{12}(p), \lambda_{13}(p), \lambda_{14}(p)$ be the four eigenvalues of $ I_1(p)$.
		Assume that  $\lambda_{11}(p)$ is the largest eigenvalue of $ I_1(p)$ and $ v(p)$ is the associated eigenvector. The edge point is thus defined as being at the point where the gradient's magnitude is at its greatest in the gradient's direction. If a direction is uniquely specified at this point \cite{tschumperle2006fast}, then the point $p$ is an edge point of the given color image where $w_1(p)$ acquires a local directional maximum in the direction $v(p)$. The foundation of this approach is the maximum of $w_1(p)$.
	\end{Method}
	
	By Theorem \ref{crlem}, we see that the spatial derivatives $\frac{\partial\mathbf{\mathbf{g}_{f_{cq}\nu}}}{\partial x_1}$ and $\frac{\partial\mathbf{\mathbf{g}_{f_{cq}\nu}}}{\partial x_2}$ can be expressed by the scale derivatives.
	The second matrix representation of the squared norm, denoted by the symbol $I_2(p)$, expresses all the spatial derivatives as the scale derivatives. Employing the eigenvalues of $ I_2(p)$, the MaSED2 method is stated as follows.
	\begin{Method} [MaSED2]\label{m4}
		The following equation  provides the generalized gradient magnitude of the MaSED2:
		\begin{equation}\label{r4}
			w_2(p)=\sum_{i=1}^{4} \lambda_{2i}(p),
		\end{equation}
		where $\lambda_{2i}(p)$, $i=1,2,3,4 $ are the four eigenvalues of the $ I_2(p)$.  This method is based on calculating the maximum of $w_2(p)$.
	\end{Method}
	
	We see that $I_1(p)$ consists of the spatial derivatives as well as the scale derivatives, and $I_2(p)$ contains only the scale derivatives. It is natural to provide a  matrix representation of the squared norm  using only the spatial derivatives by Theorem \ref{crlem}. We denote this matrix representation  by $I_3(p)$ (refer to Appendix \ref{a_sec4} for  details), then the following method is obtained. 
	
	\begin{Method} [MaSED3]\label{m5}
		Let $ \lambda_{3i}(p)$ be the four eigenvalues of $ I_3(p)$,  $ i=1,2,3,4.$
		The {generalized gradient magnitude } of the MaSED3 is given by   
		\begin{equation}\label{r5}
			w_3(p)=\sum_{i=1}^{4} \lambda_{3i}(p).
		\end{equation}
		This method is based on calculating the maximum of $w_3(p)$.
	\end{Method}

	\section{Experiments}\label{hxx4}

	\subsection{Algorithms }
	\begin{description}
		\item [Step 1.]
		Input the color image $f=f_{1}{\e_1}+f_{2}{\e_2} +f_3{\e_3}$.
		\item [Step 2.]  Using  Eq. (\ref{A}) to perform Poisson filtering and  conjugate Poisson filtering on $f(x_1,x_2)$,
		we obtain the CCHS: $ f_{cq}(z_{1},z_{2} )$.
		\item [Step 3.]
		For a given color $\nu =a{\e_1}+b{\e_2}+c{\e_3}$,  compute the $ {\rm{Sc}}[f_{cq}\nu]$ and $ \theta$ as in Eqs. (\ref{real})
		and (\ref{phase}) respectively.
		\item [Step 4.]
		Select  an edge detection method.
		\par
		CHED   method (The scales $ y_{1}=y_{2}=2$).
		\begin{itemize}
			\item  Compute the derivatives of    $ {\rm{Sc}}[f_{cq}\nu]$ and $ \theta$  with respect to $x_1$ and $x_2$, respectively.
			\item  Compute $ \lambda_{+} $ as in Eq. (\ref{r1}).
		\end{itemize}
		
		MCHED   method (The scales $ y_{1}=y_{2}=8$).
		\begin{itemize}
			\item  Compute $B_1,B_2,D_1,D_2$ (see Appendix \ref{a_sec2}), respectively.
			\item  Compute $\widetilde{\lambda}_{+} $ as in Eq. (\ref{r2}).
		\end{itemize}
		
		MaSED1 method (The scales $ y_{1}=y_{2}=2$).
		\begin{itemize}
			\item Compute the derivatives of   $ {\rm{Sc}}[f_{cq}\nu]$ and $ \theta$  with respect to $x_1$ and $x_2$, and the scale derivatives of 
			$ {\rm{Sc}}[f_{cq}\nu]$ and $ \theta$  with respect to $y_1$ and $y_2$.
			\item  Compute the first matrix representation of the metric $ (d\mathbf{\mathbf{g}_{f_{cq}\nu}})^2$, i.e., $I_1(p)$.
			\item  Compute $ w_1(p) $ as in Eq. (\ref{r3}).
		\end{itemize}
		
		MaSED2 method (The scales $ y_{1}=y_{2}=2$).
		\begin{itemize}
			\item  Compute the second matrix representation of the metric $ (d\mathbf{\mathbf{g}_{f_{cq}\nu}})^2$, i.e., $I_2(p)$.
			\item  Compute $ w_2(p) $ as in Eq. (\ref{r4}).
		\end{itemize}
		
		MaSED3 method (The scales $ y_{1}=y_{2}=2$).
		\begin{itemize}
			\item  Compute the third  matrix representation of the metric $ (d\mathbf{\mathbf{g}_{f_{cq}\nu}})^2$, i.e., $I_3(p)$.
			\item  Compute $ w_3(p) $ as in Eq. (\ref{r5}).
		\end{itemize}

		\item[Step 5.]
		Steps 1 through 4 are completed to produce the gradient map. We perform non-maximum suppression operation (with radius $r=1.5$) to the gradient map for each test image to make the edges narrower.
	\end{description}

	\subsection{Edge detection performance evaluation in relation to a specific color}
	To look into how well various edge detectors perform. In order to assess the accuracy and noise robustness of edge detection, edge detection operators' performance is measured quantitatively and qualitatively. The quantitative metrics used in this section  are the Peak Signal-to-Noise Ratio  ({PSNR}), the Structural Similarity Index Measure  ({SSIM}) \cite{wang2004image},  the Feature Similarity  Index Measure ({FSIM}) \cite{zhang2011fsim}, and the standard measure \textbf{F} \cite{abdou1979quantitative}.

	The qualitative measure  is also highly significant in image processing \cite{plataniotis2000color,androutsos1998color}.
	The following criteria are utilized for subjective judgment.
	The first one is edge continuity; strong edge continuity is preferred to broken edge continuity in edge maps.
	The second is the thinness of the edges; in edge maps, thin lines indicate that the edge operator has a high degree of certainty when determining the edges. The third one is the ability to reduce noise, and the final one is visual appeal.

	In this part, we compare our approaches with four existing approaches for edge detection, including   K-means clustering method, nearest neighbor approach, and the color monogenic signal-based approaches, i.e., Color Metric Edge Detection (CMED) and Color Metric and Multiscale Edge Detection (CMMED) \cite{demarcq2011color}. 
	
\begin{figure*}[!ht]
					\centering
					\subfloat[]{\includegraphics[width=2.26in]{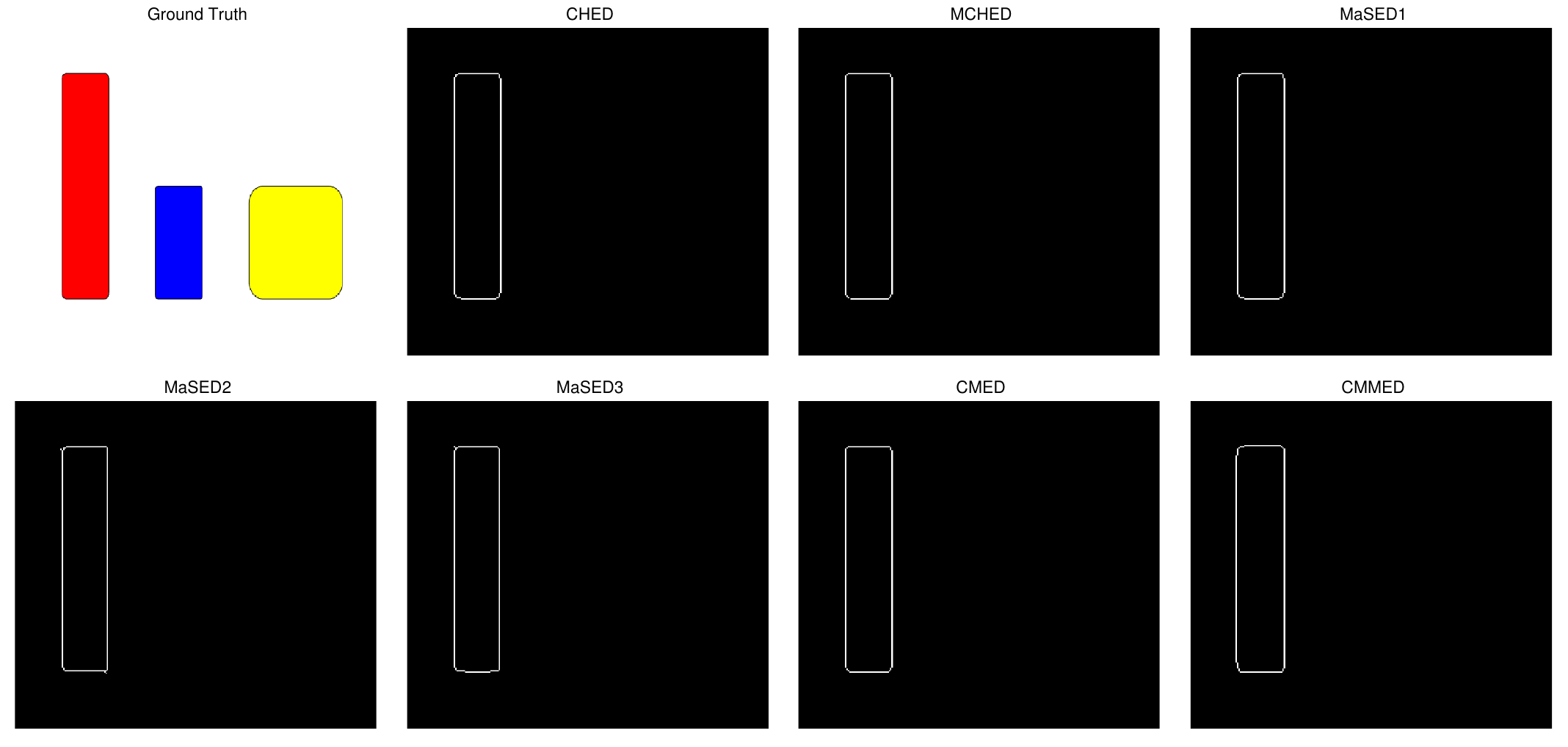}%
					}
					\hfil
					\subfloat[]{\includegraphics[width=2.26in]{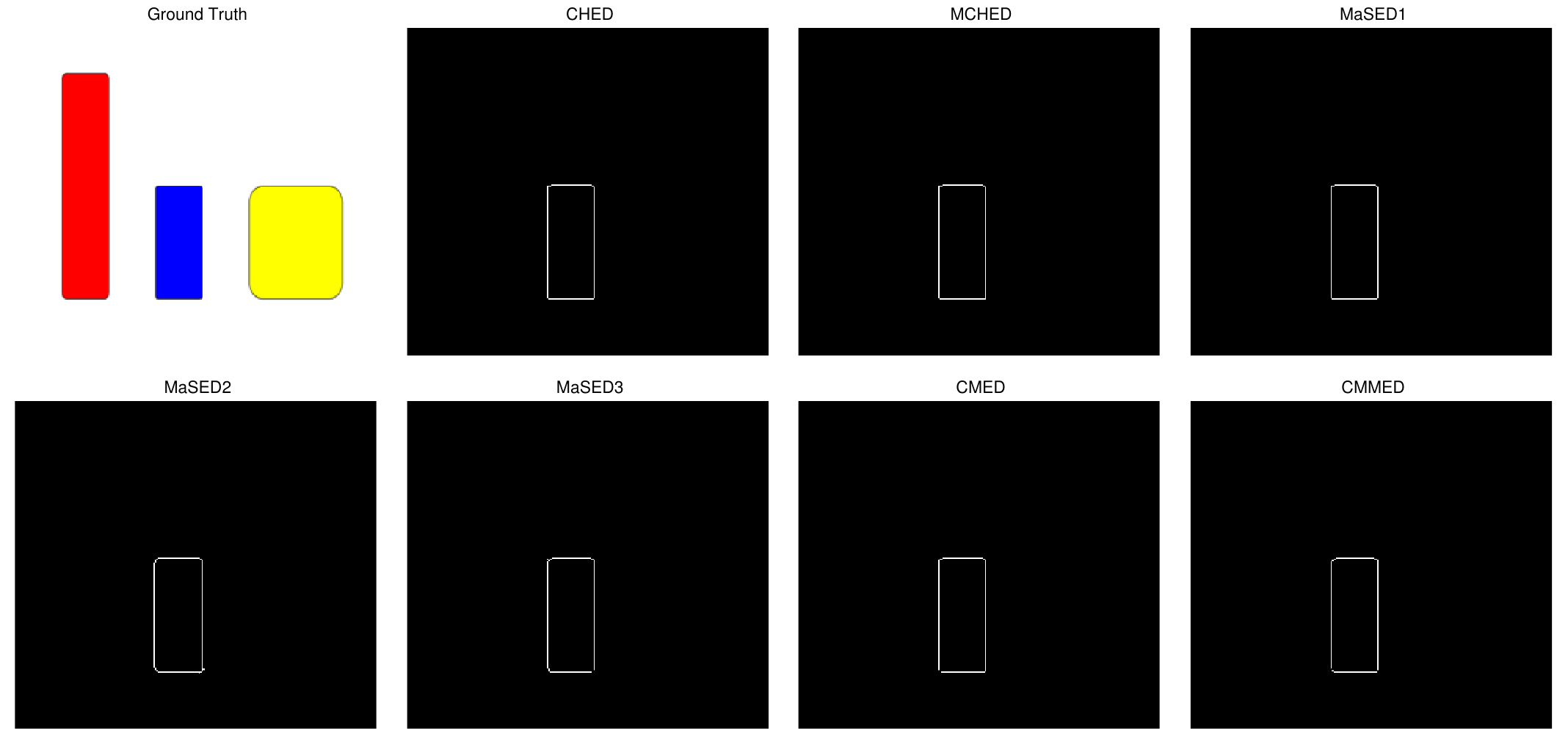}%
					}
					\hfil
					\subfloat[]{\includegraphics[width=2.26in]{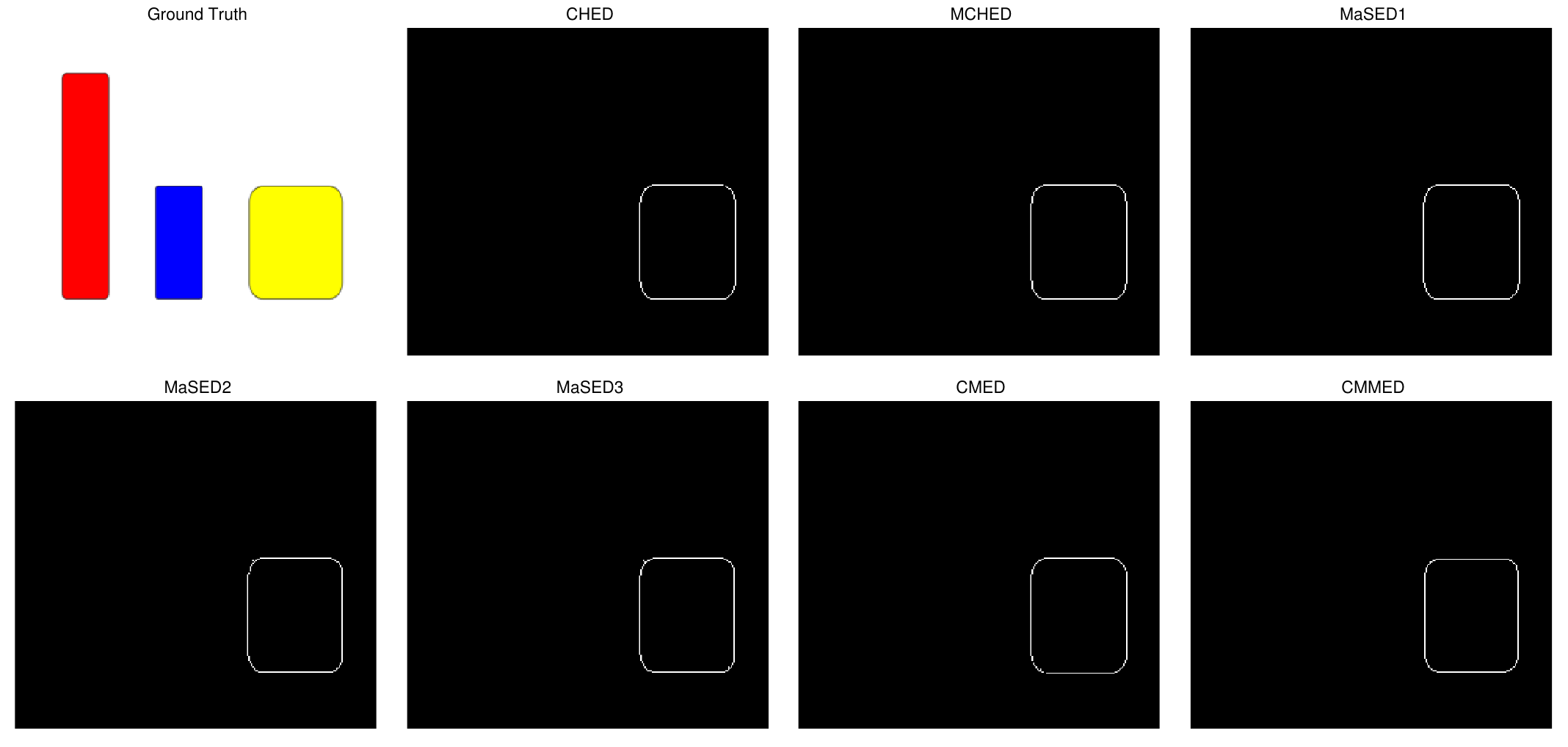}%
					}
					\caption{Results of   different algorithms to detect the edge points of the rectangles.}
					\label{rectanges}
				\end{figure*}

	\begin{example}[The precision of edge operators in identifying the borders of a synthetic image]  
		The effectiveness of the edge operators in terms of the precision of edge detection is assessed using a simulated image. The synthetic image's edge information is known. Three rectangles are present, each with a different color and curvature: the blue rectangle has a curvature of 0.2, the red rectangle has a curvature of 0.2, and the yellow rectangle has a curvature of 0.1. The quantitative values of   accuracy o  are listed in Table \ref{tab2} and the detection results are presented in Figure \ref{rectanges}. It   demonstrates that MaSED1, MCHED, and CHED produce superior outcomes to the other edge operators. As shown in Table  \ref{tab2}, CHED and MaSED1 perform similarly and all provide high-quality edge maps for noiseless images.
	\end{example}
	
\begin{figure*}[!ht]
					\centering
					\subfloat[]{\includegraphics[width=2.26in]{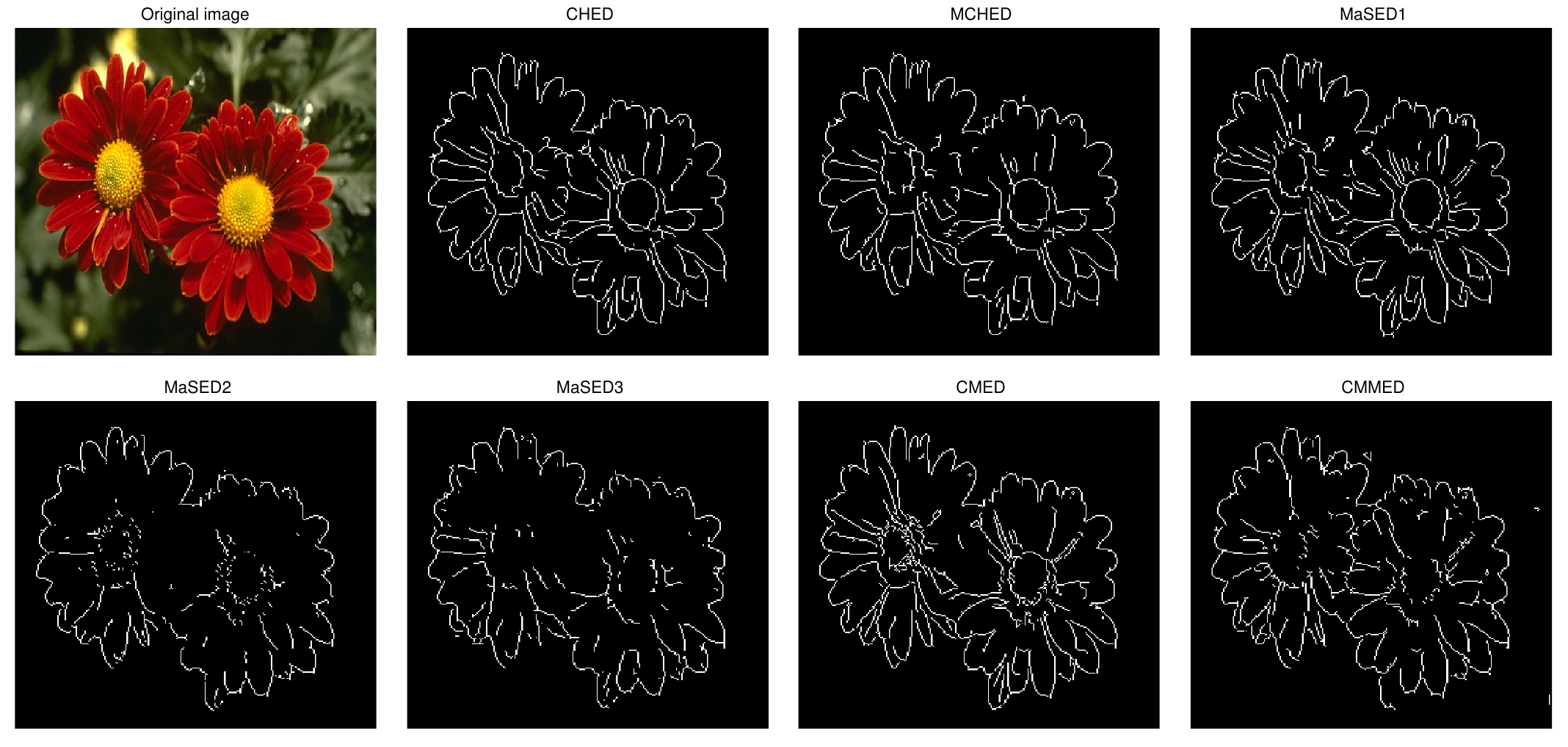}%
					}
					\hfil
					\subfloat[]{\includegraphics[width=2.26in]{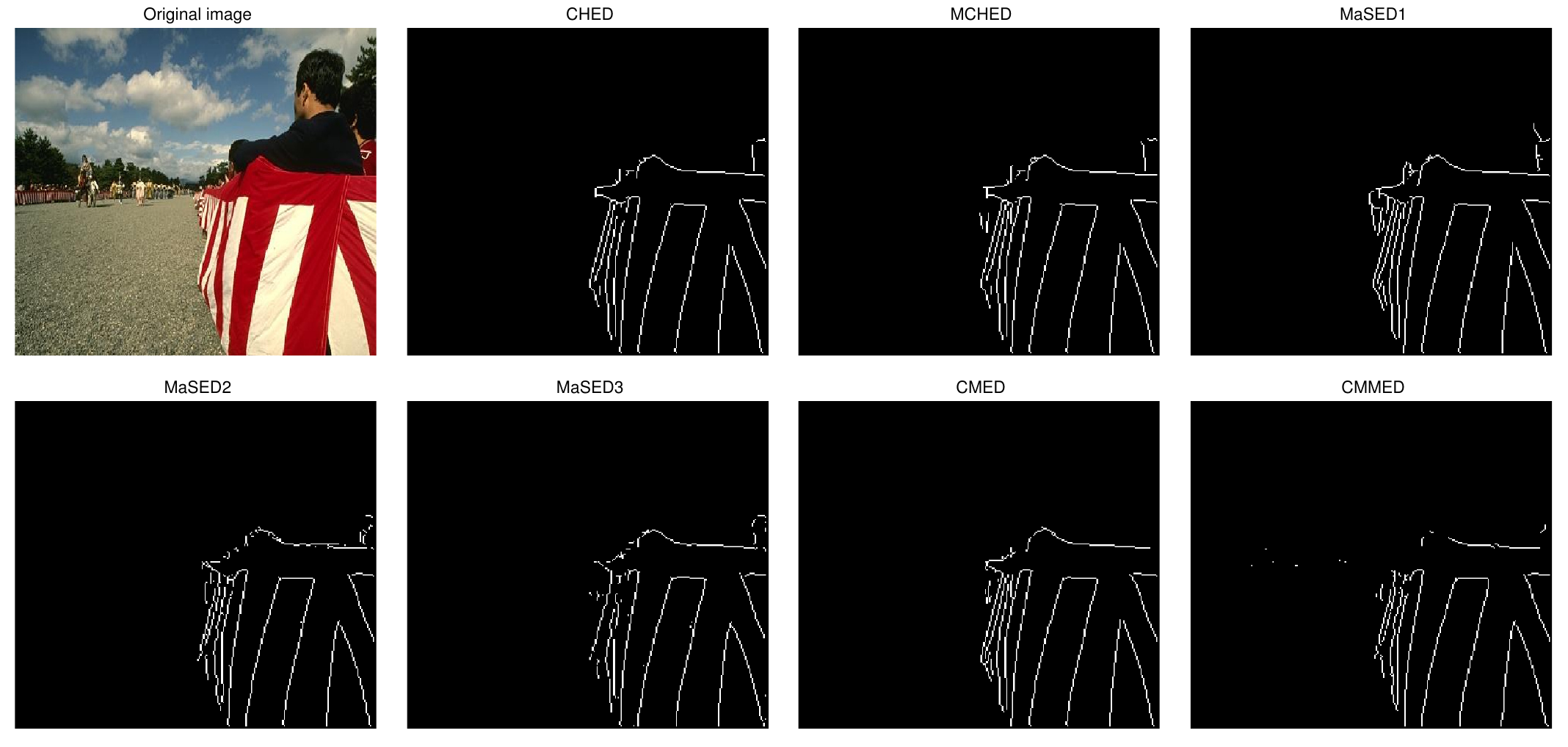}%
					}
					\hfil
					\subfloat[]{\includegraphics[width=2.26in]{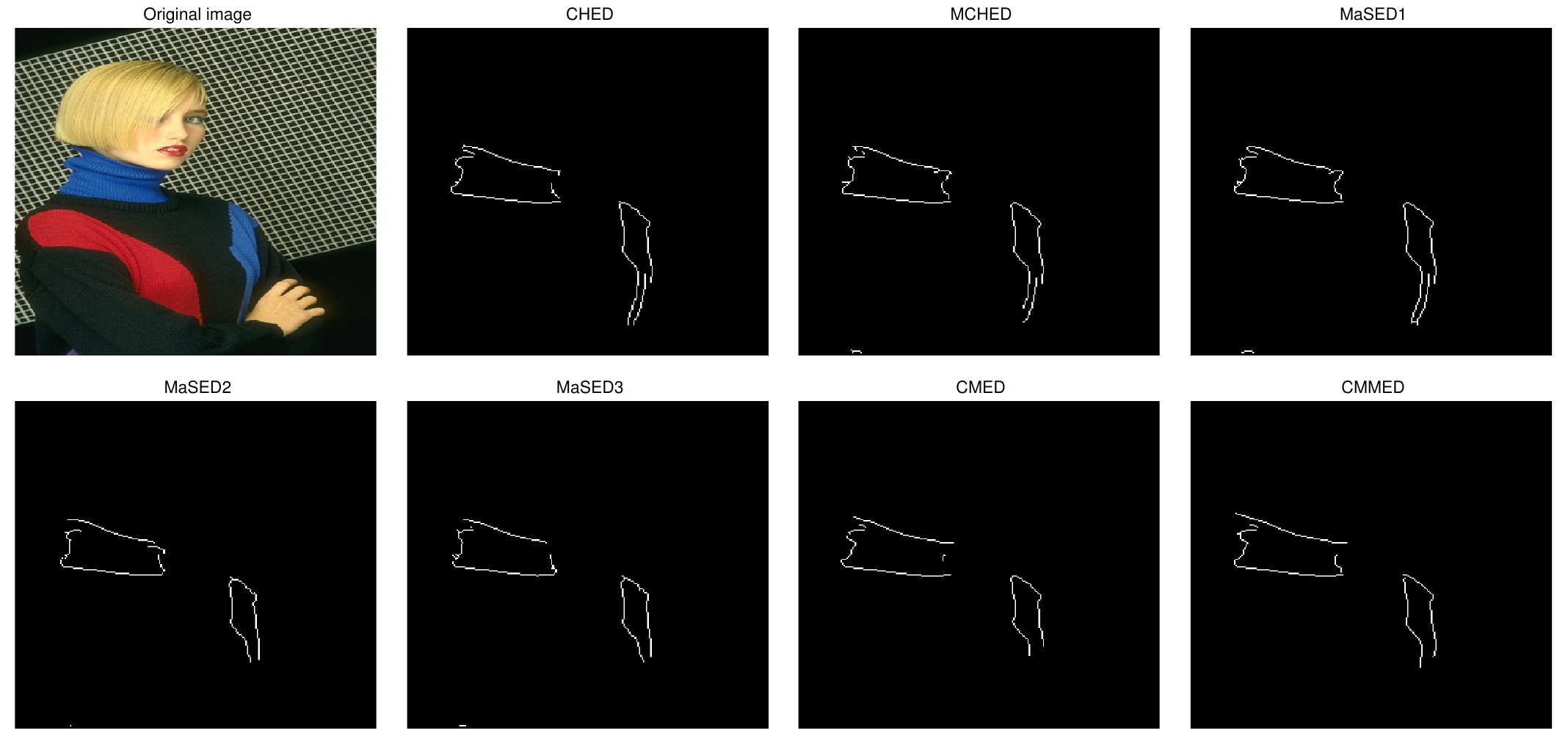}%
					}\\
					\subfloat[]{\includegraphics[width=2.26in]{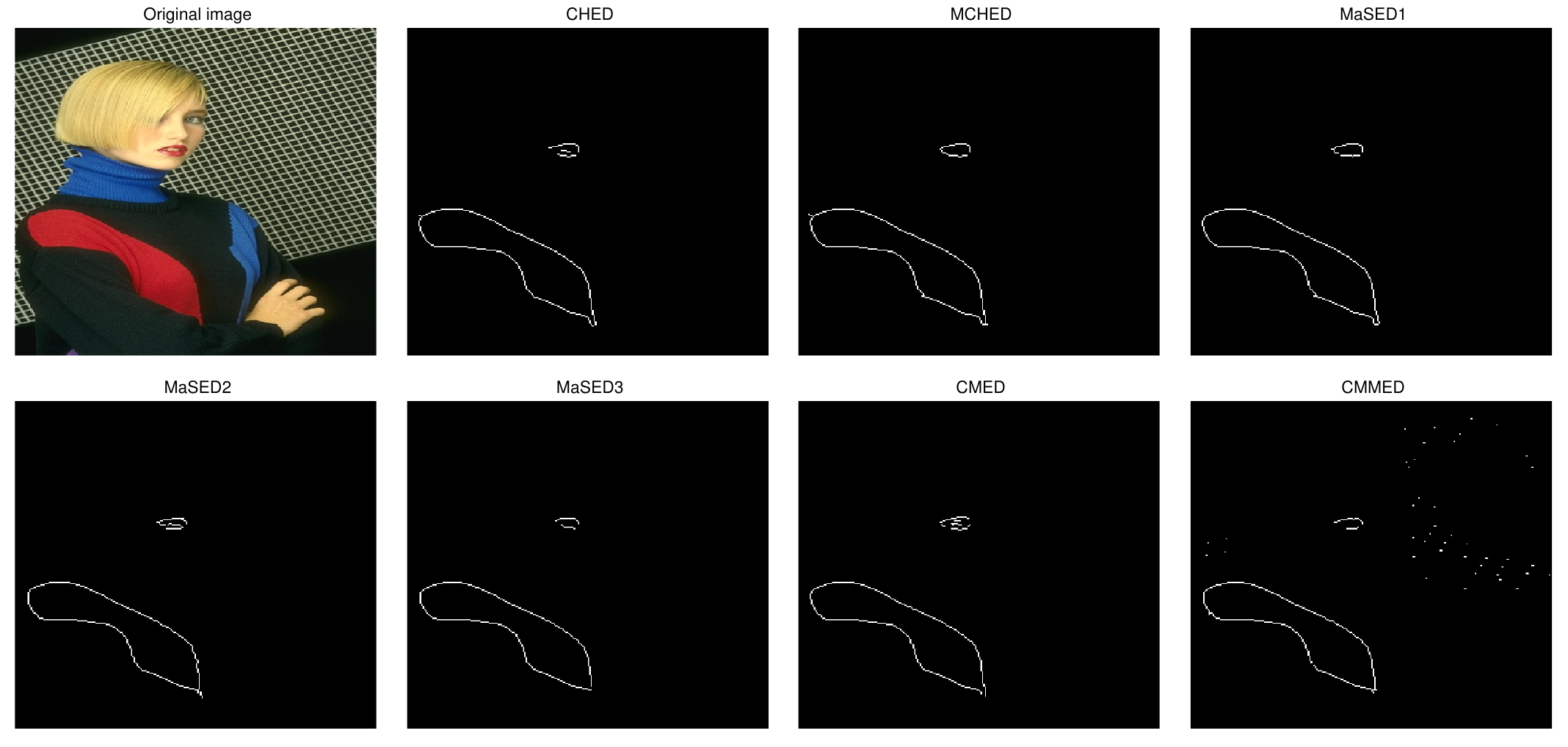}%
					}
					\hfil
					\subfloat[]{\includegraphics[width=2.26in]{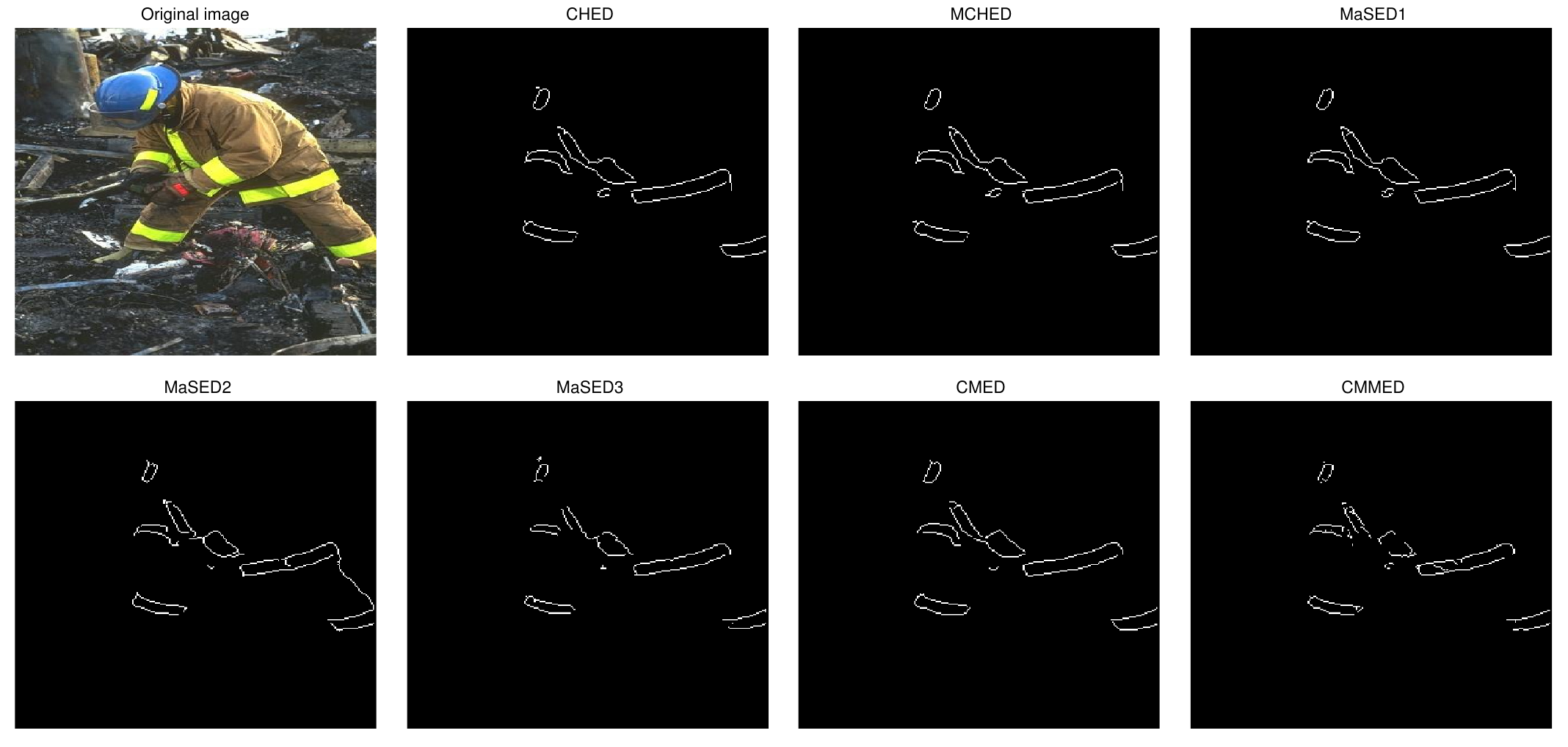}%
					}
					\hfil
					\subfloat[]{\includegraphics[width=2.26in]{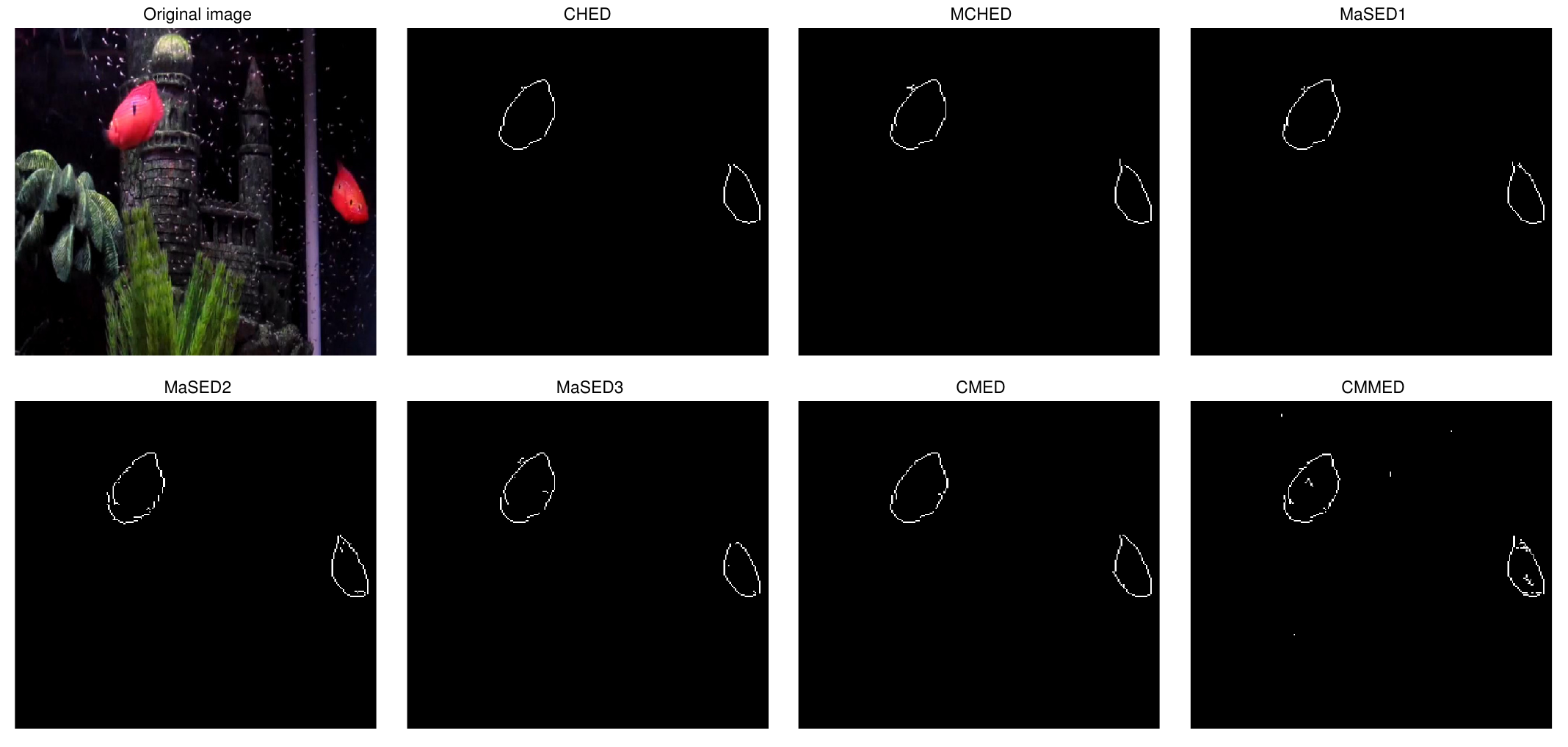}%
					}
					\caption{Results of   different algorithms to detect the edge points of the real-world  images.}
					\label{realimage}
				\end{figure*}

	
	\begin{example}[The validity of the edge operators for real-world images]
		The real-world images in Figure \ref{realimage} are used for this experiment. They are \emph{Flowers},  \emph{Woman}, \emph{Fireman},   \emph{Fence} and \emph{Fish} from 
		the Berkeley image dataset \cite{martin2001database}.
		\begin{enumerate} 
			\item
			In subfigure (a), we employ various edge detection techniques to "Flowers". With the exception of MaSED2 and MaSED3, all approaches can distinguish the flowers from the background and segment the yellow and red sections.
			\item
			More edge points of the red-colored portions of the "Fence" are detected by MaSED1 than by the other edge operators.
			\item
			The edge points of the blue sections of "Woman" are accurately detected by MaSED1, CHED, and MCHED. However, the only continuous edge map is the one produced by MaSED1.
			\item
			All edge operators are able to identify the edge points of the red portions of "Woman". However, CMMED's performance is quite subpar.
			\item
			The green reflective strips on Man's firefighter suit had more edge points detected by MaSED1, CHED, and MCHED than the others. MaSED1, CHED, and MCHED outperform the others in terms of aesthetic appeal, according to \cite{androutsos1998color}.
			\item
			The edge operators are able to identify the red fish's edge points. CHED provides edges that are more continuous and thinner than the competitors. Additionally, MaSED2, MaSED3, and CMMED all perform fairly similar.
		\end{enumerate}		
	\end{example}

\begin{figure*}[!ht]
					\centering
					\subfloat[]{\includegraphics[width=2.26in]{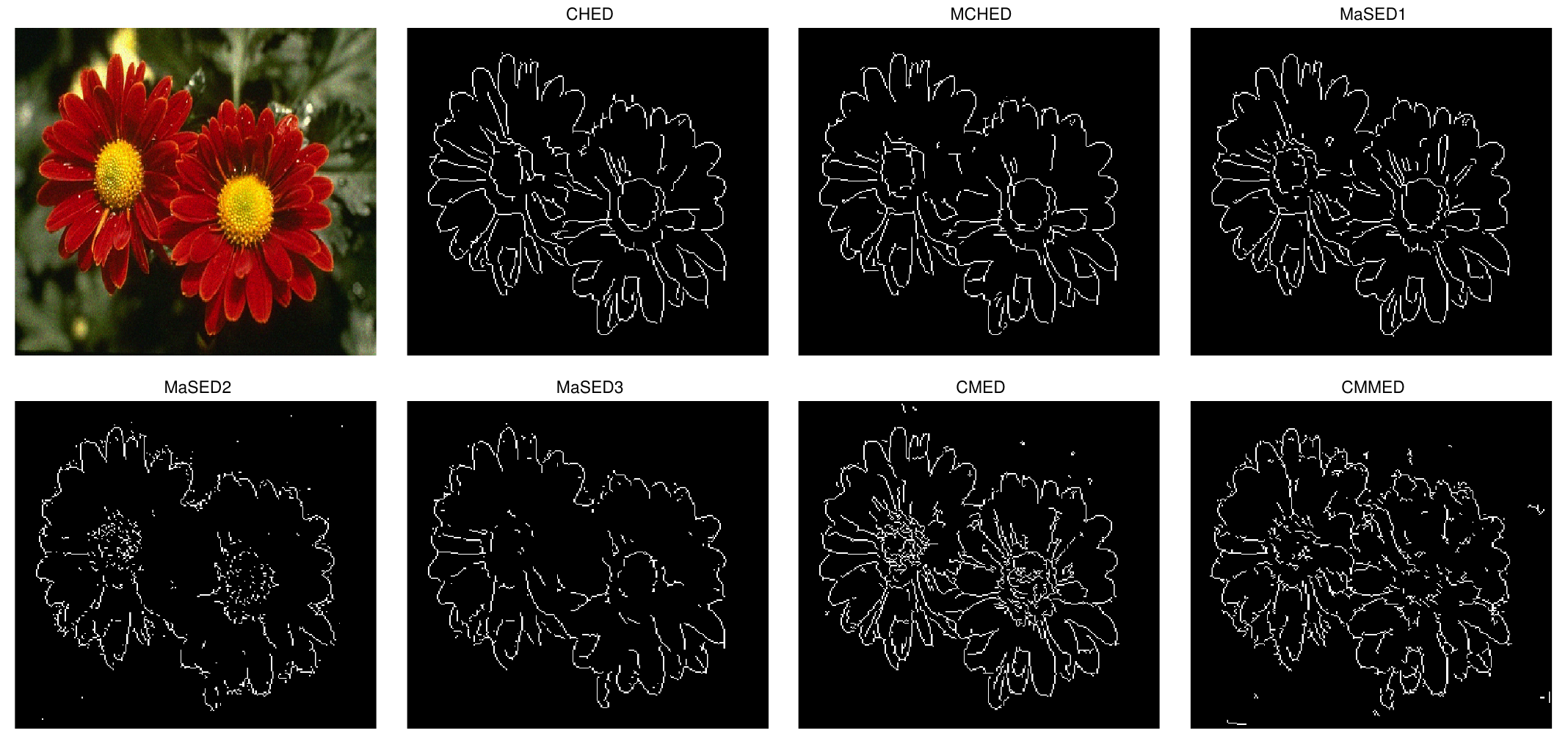}%
					}
					\hfil
					\subfloat[]{\includegraphics[width=2.26in]{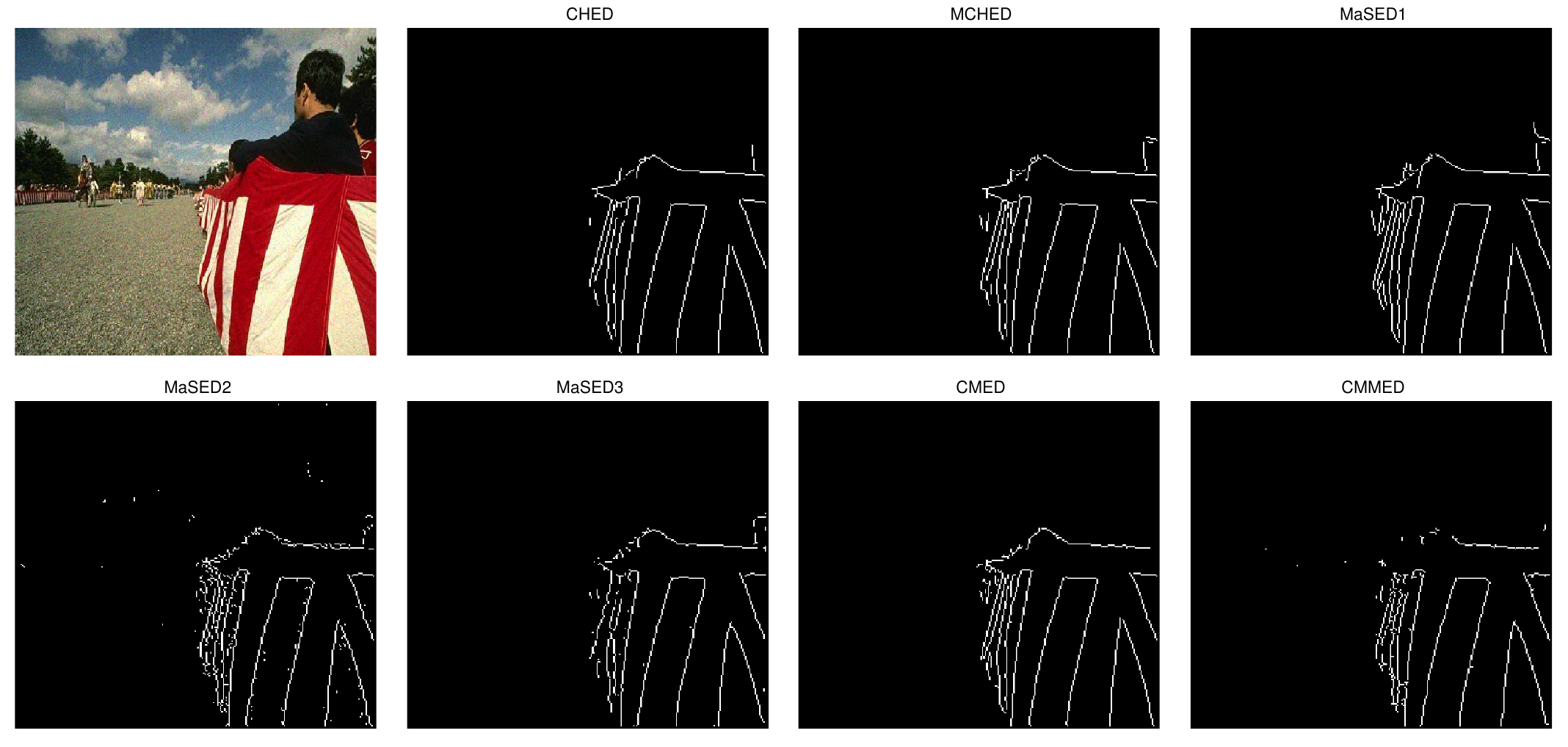}%
					}
					\hfil
					\subfloat[]{\includegraphics[width=2.26in]{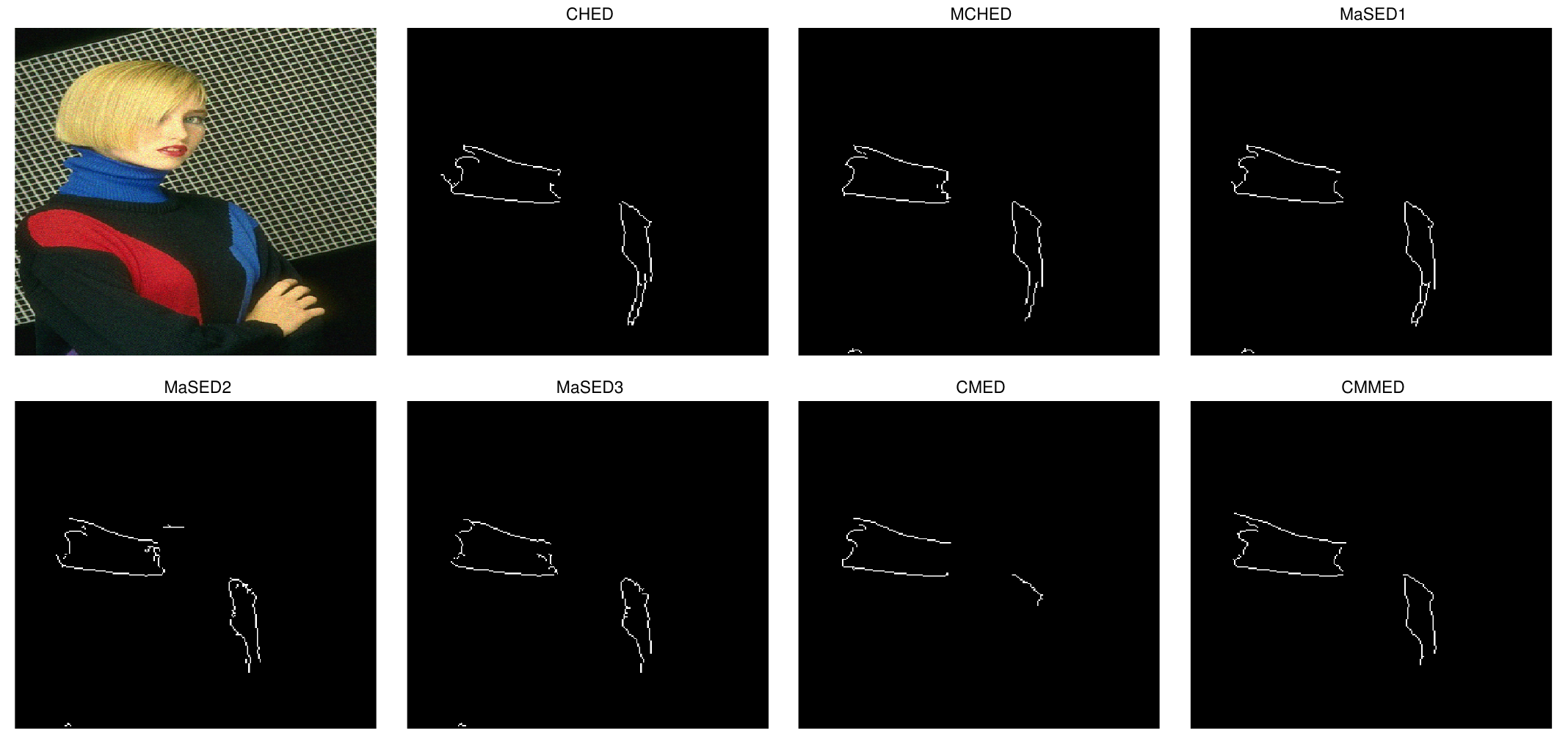}%
					}\\
					\subfloat[]{\includegraphics[width=2.26in]{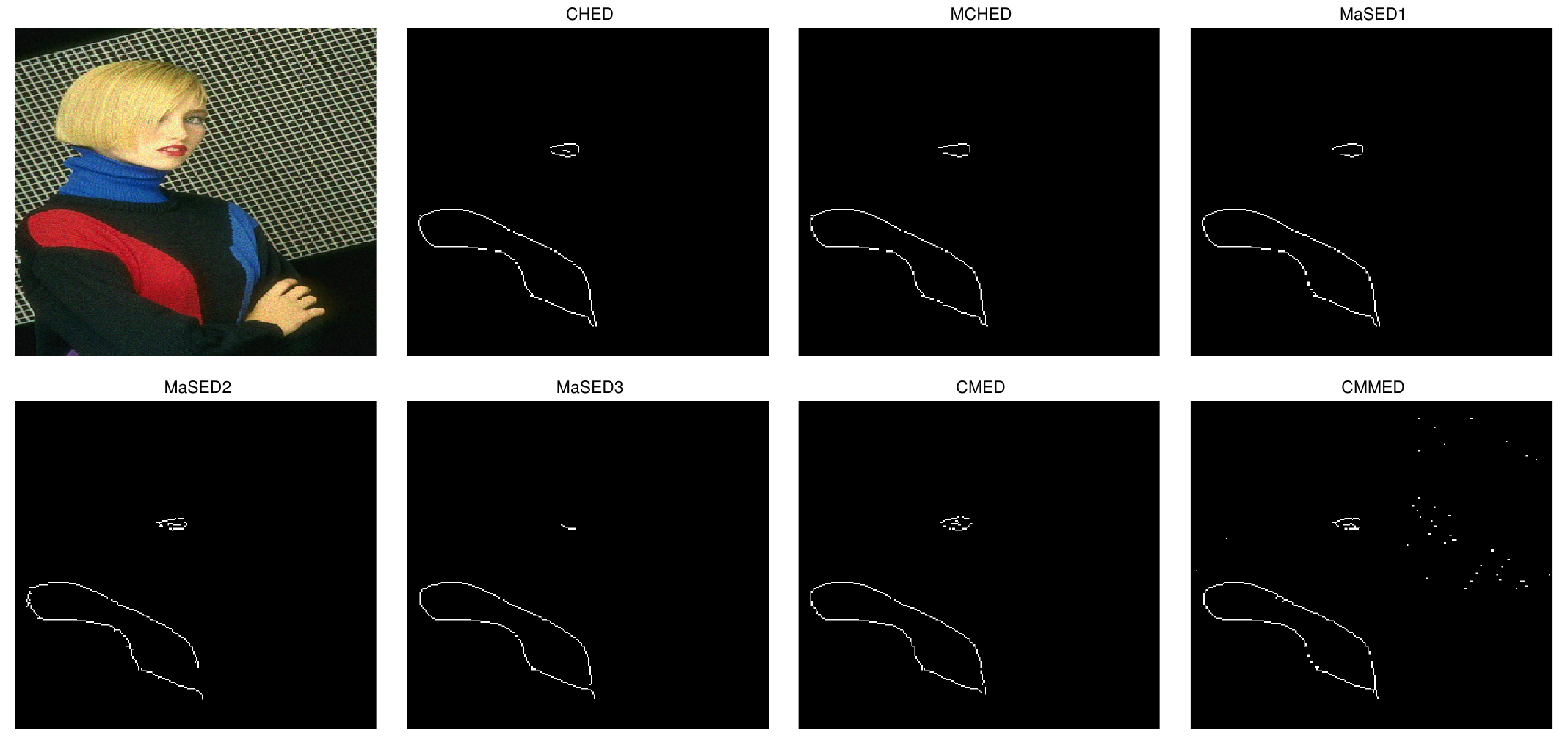}%
					}
					\hfil
					\subfloat[]{\includegraphics[width=2.26in]{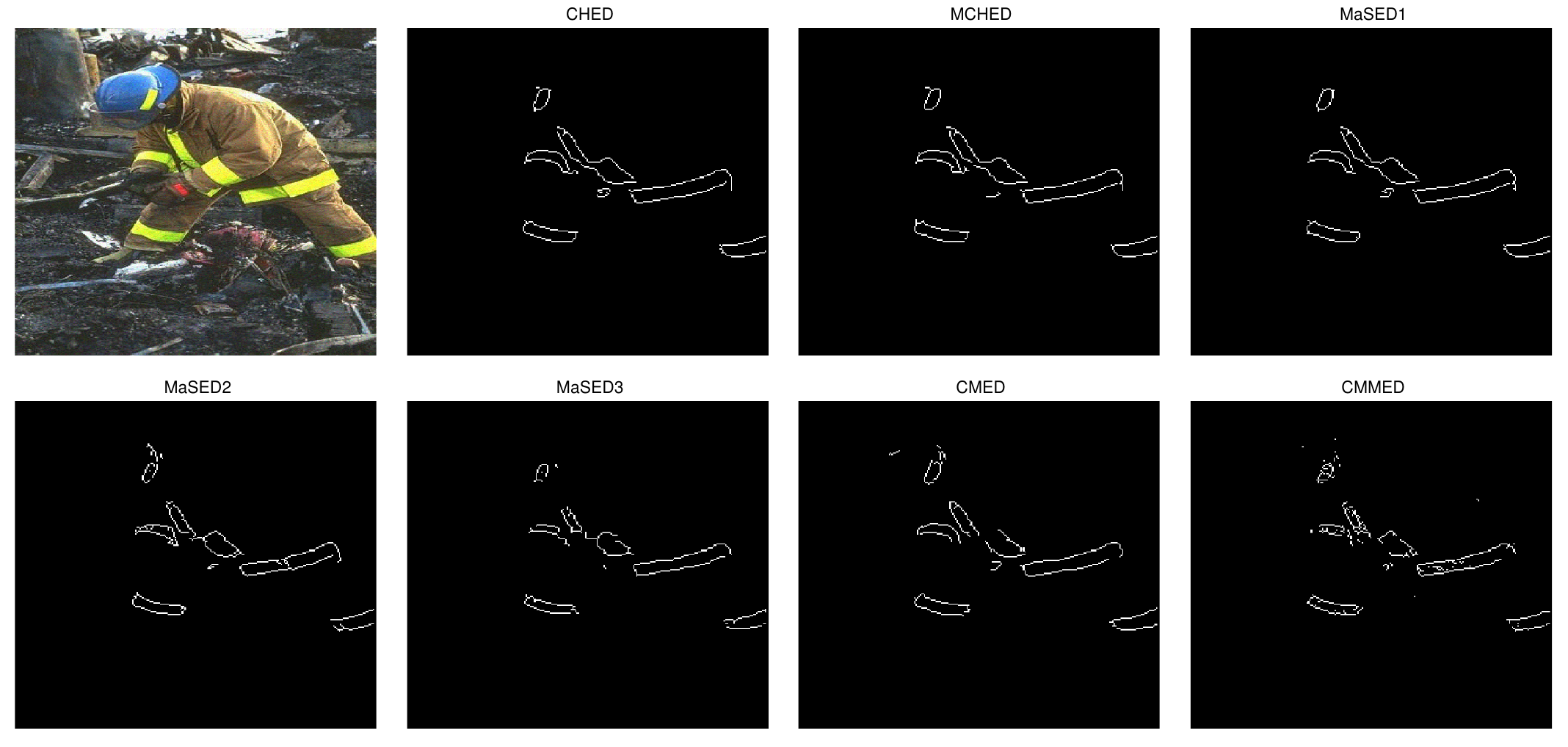}%
					}
					\hfil
					\subfloat[]{\includegraphics[width=2.26in]{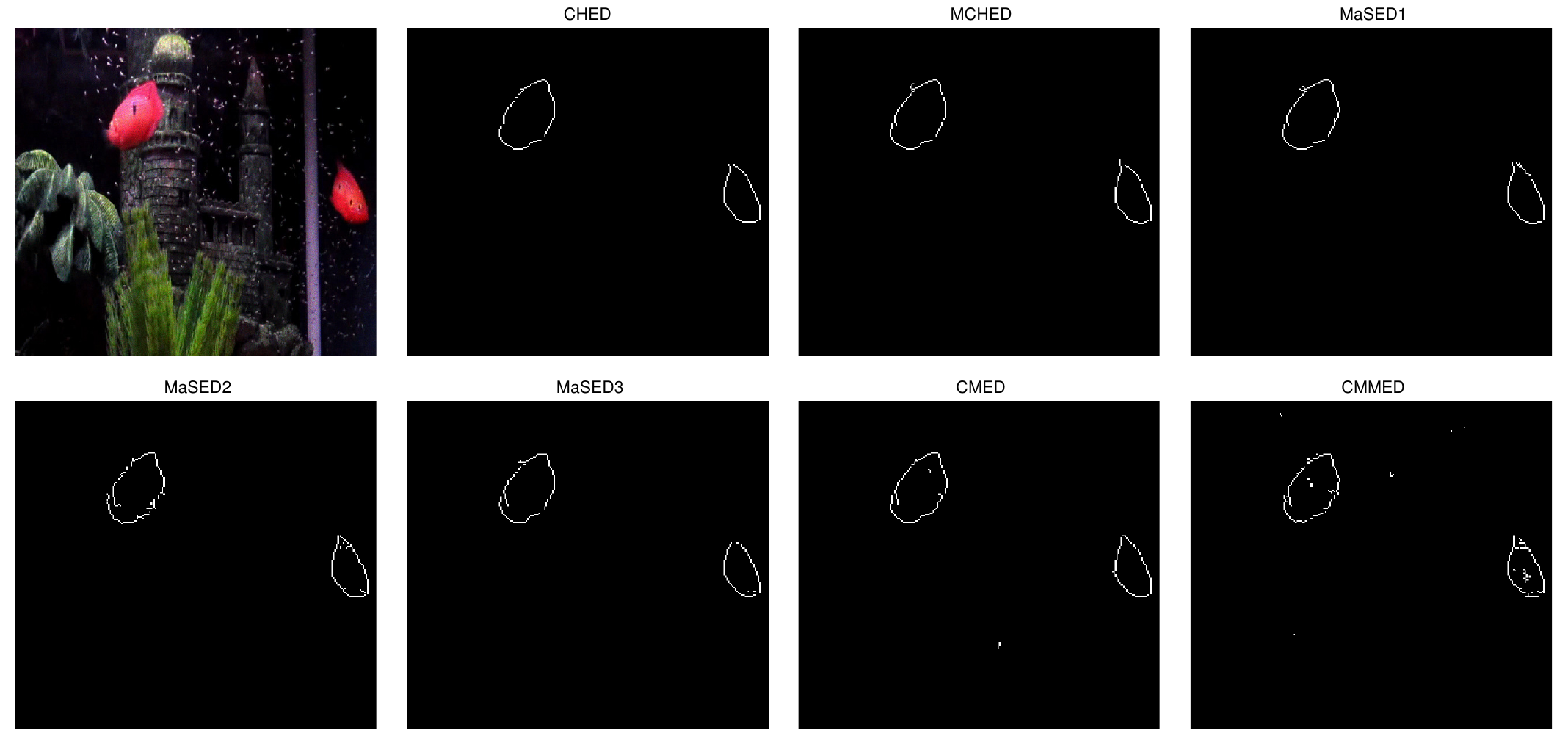}%
					}
					\caption{Results of   different algorithms to detect the edge points of the real-world  images corrupted by   Poisson noise.}
					\label{realimagep}
				\end{figure*}

\begin{example}[Edge detection for real-world images with Poisson noise]
For this experiment, the real-world images that have been tainted by Poisson noise are used. We can observe from a comparison of Figures \ref{realimage} and \ref{realimagep} that CHED, MCHED, and MaSED1 perform well on edge detection even when Poisson noise is present. Table \ref{tab4} demonstrates that MaSED1, MCHED, and CHED are more noise-resistant than the rest. Furthermore, the visual evaluations concur with the qualitative measurements as well. By raising the scale values, CMMED and CMED can perform their anti-noise ability better.
\end{example}

\begin{figure*}[!htb]
					\centering
					\includegraphics[height=4cm,width=16.5cm]{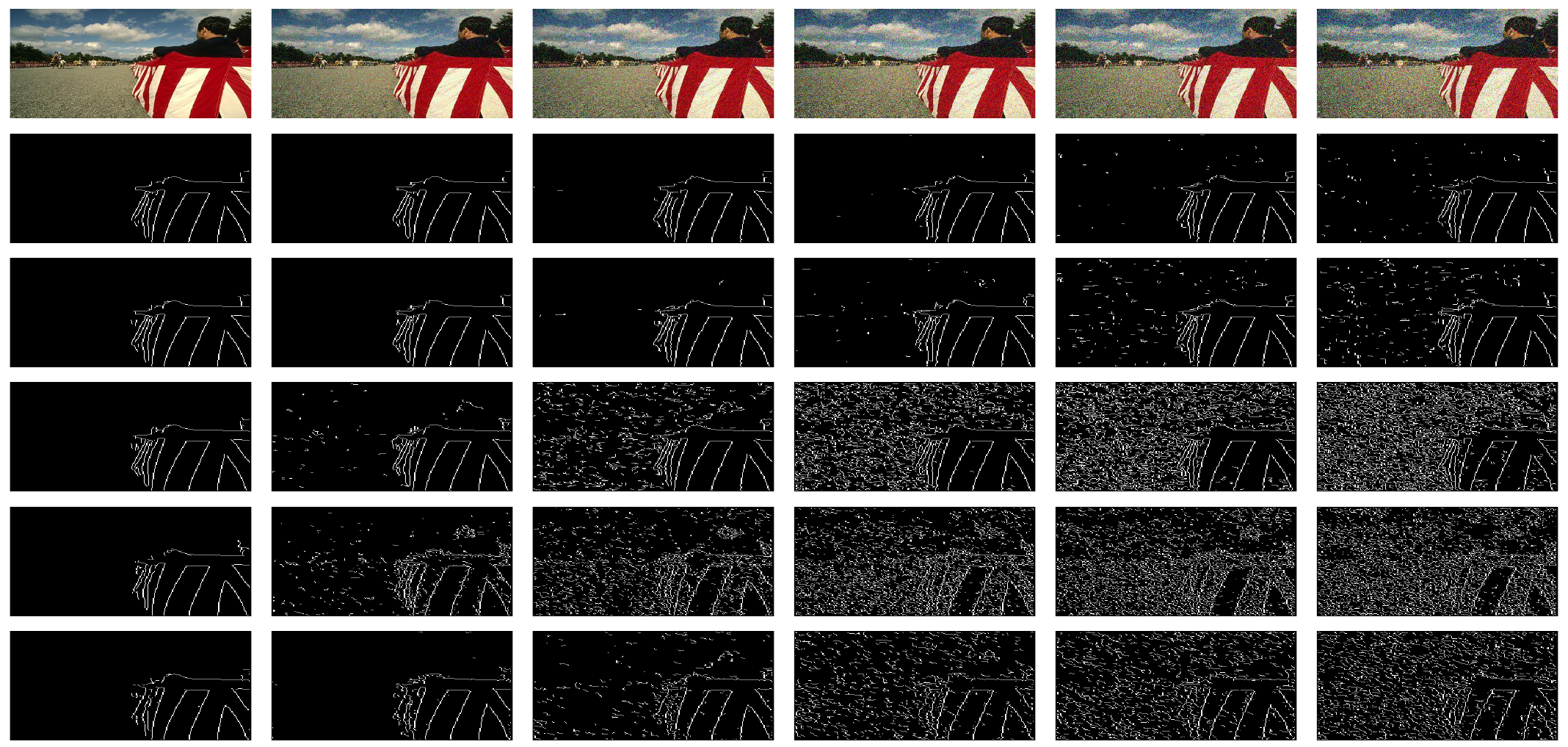}
					\caption{Edge detection of  red  color in the presence of Gaussian noise.  Row 2--6 correspond to     CHED,   MCHED,   MaSED1,  MaSED2 and  MaSED3, respectively. From  left to right: the original image and the image corrupted by Gaussian noise with  variance from $0.01$ to $0.05$.}
					\label{redg1}
				\end{figure*}

\begin{figure*}[!htb]
					\centering
					\includegraphics[height=4cm,width=16.5cm]{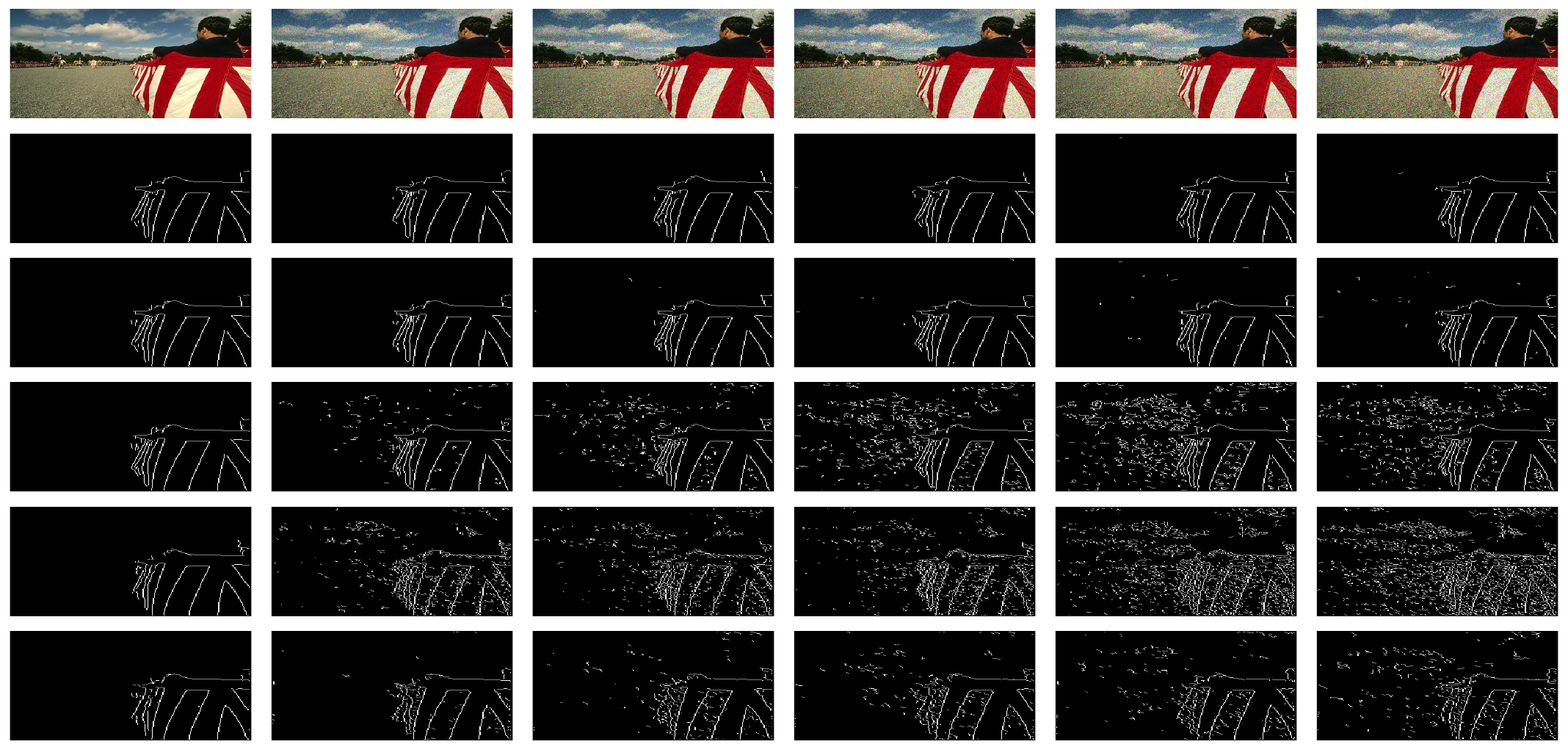}
					\caption{Edge detection of  red  color in the presence of speckle  noise.  Row 2--6 correspond to     CHED,   MCHED,   MaSED1,  MaSED2 and  MaSED3, respectively. From  left to right: the original image and the image corrupted by speckle  noise with  variance from $0.05$ to $0.09$.}\label{redsp1}
				\end{figure*}

\begin{figure*}[!htb]
					\centering
					\includegraphics[height=4cm,width=16.5cm]{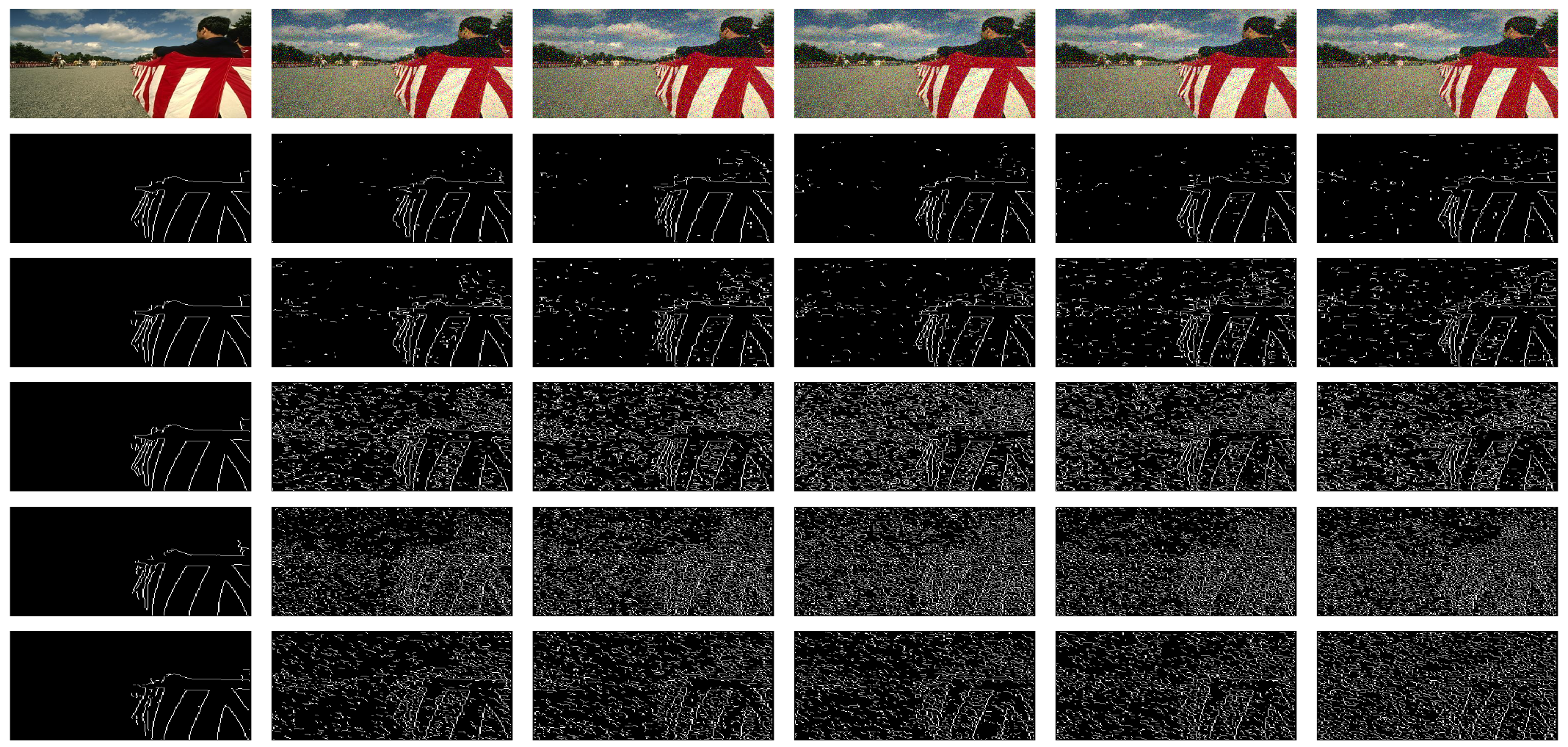}
					\caption{Edge detection of  red  color in the presence of salt and pepper   noise.  Row 2--6 correspond to     CHED,   MCHED,   MaSED1,  MaSED2 and  MaSED3, respectively. From  left to right: the original image and the image corrupted by salt and pepper  noise with  density  from    $0.10$ to $0.15$.} 
					\label{redst1}
				\end{figure*}

\begin{figure*}[!htb]
					\centering
					\includegraphics[height=4cm, width=12cm]{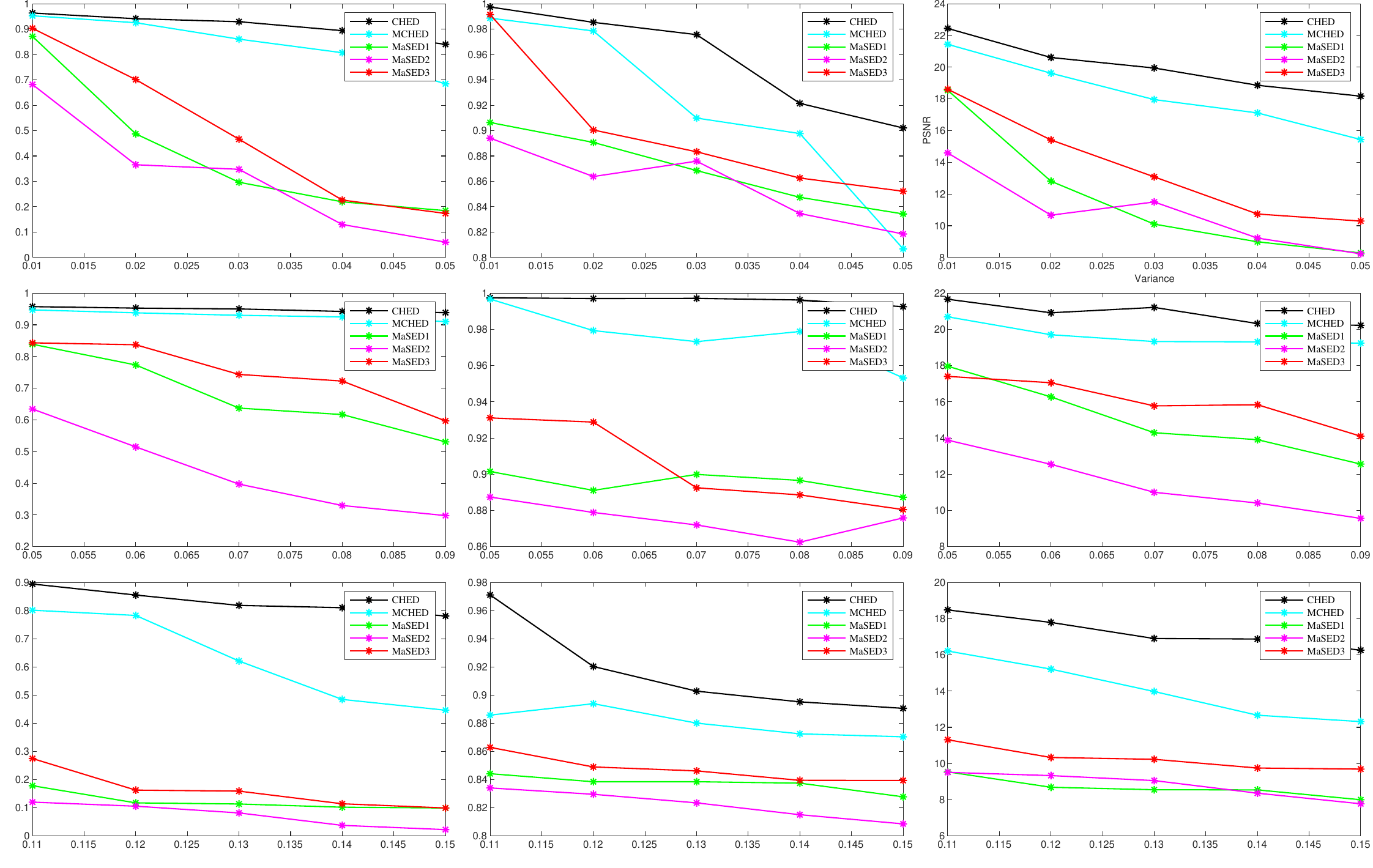}
					\caption{ From left to right:  SSIM, FSIM and PSNR values of  edge detection on noisy images.   Row 1--3 correspond to  the edge detection results in the presence of   Gaussian noise,  speckle noise, salt and pepper noise,  respectively.}
					\label{rednoise}
				\end{figure*}

	\begin{example}[Edge detection in various variances and intensities of noise]
		Gaussian noise with variances ranging from $0.01$ to $0.05$, speckle noise with variances ranging from $0.05$ to $0.09$, and salt and pepper noise with densities ranging from $0.01$ to $0.05$, all consecutively, taint the fence image.
		Figures \ref{redg1},  \ref{redsp1}, and \ref{redst1} display the edge detection outcomes obtained using our methodologies. We can observe that MaSED3, MCHED, and CHED produce the best results. The CHED approach yields a very good result for pinpointing the margins of the red portions of the image that have been tainted by salt and pepper noise.
		The SSIM, FSIM, and PSNR values for various approaches are shown in Figure  \ref{rednoise}. Evidence suggests that CHED performs at its peak when there is a lot of noise.
	\end{example}

\begin{figure*}[!htb]
					\centering
					\includegraphics[height=4cm, width=12cm]{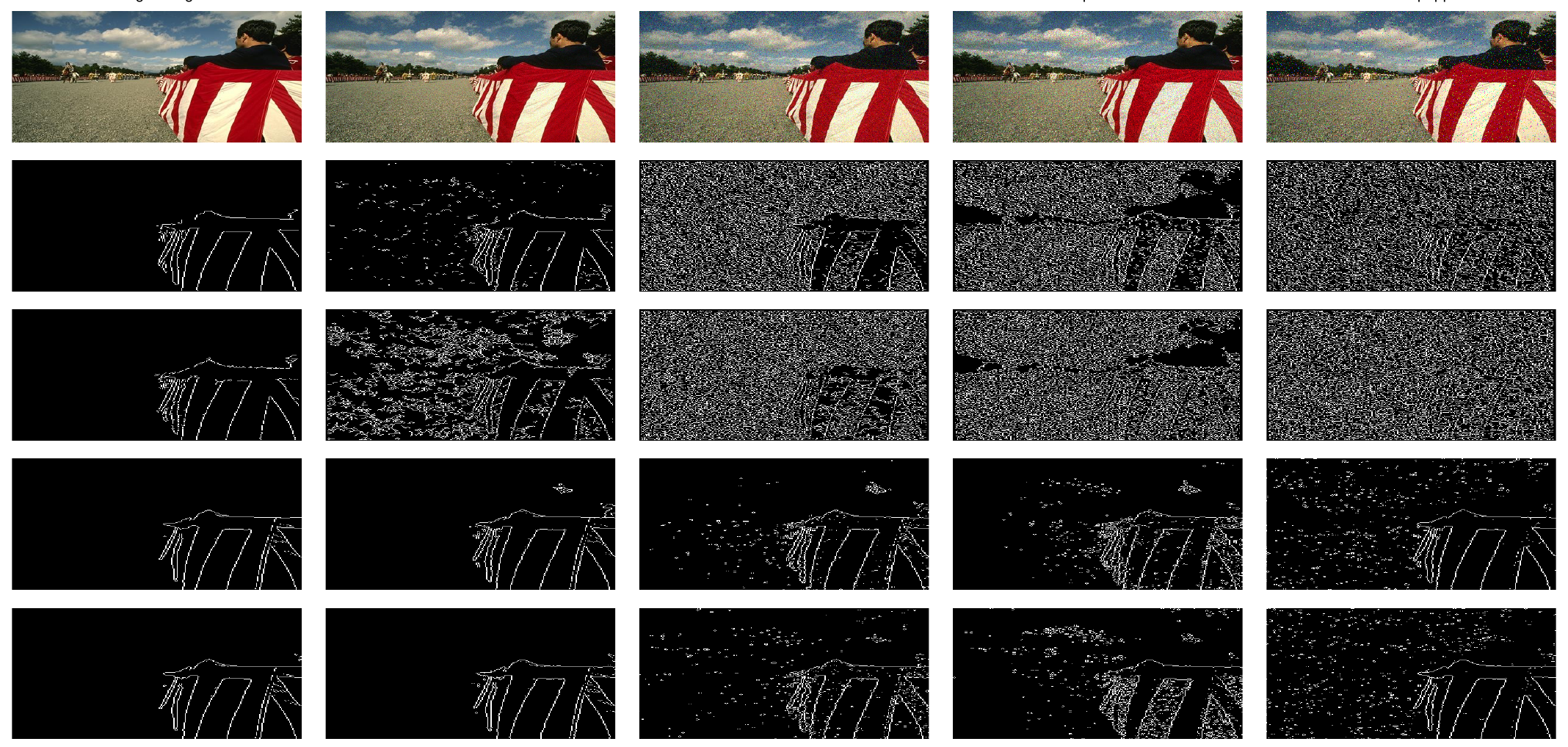}
					\caption{The first row shows the original image and the images  corrupted by Poisson, Gaussian, speckle and salt and pepper noises, respectively. Row 2--5 correspond to  the edge detection results  of   CMED \cite{demarcq2011color}, CMMED \cite{demarcq2011color}, K-means clustering method \cite{arthur2007k} and nearest  neighbor method \cite{laaksonen1996classification}.}
					\label{cmed}
				\end{figure*}

	\begin{example}[Comparative study with  other methods]
		We compare our approaches with those of K-means clustering,  nearest neighbor, CMED, and CMMED \cite{demarcq2011color}. The edges of a specific color are obtained by applying the clustering and Sobel edge operator successively for K-means clustering and nearest neighbor algorithms.
		Poisson noise, Gaussian noise with a variance of $0.01$, speckle noise with a variance of $0.05$, and salt and pepper noise with a density of $0.01$ are used to distort the image. The detection outcomes of the nearest neighbor method (NNM), CMED, and CMMED algorithms are shown in Figure \ref{cmed}. As shown in Figures \ref{realimagep}, \ref{redg1}, \ref{redsp1} and  \ref{redst1}, our approaches successfully detect the edges of noisy pictures. However, in the presence of noise, the performance of CMED and CMMED approaches is not as desirable. Table \ref{other} demonstrates that our methods outperform the competition.
		
	\end{example}

	\subsection{Application to optical flow}\label{hxxsflow}
	
\begin{figure*}[!htb]
					\centering
					\includegraphics[height=3cm,width=16.5cm]{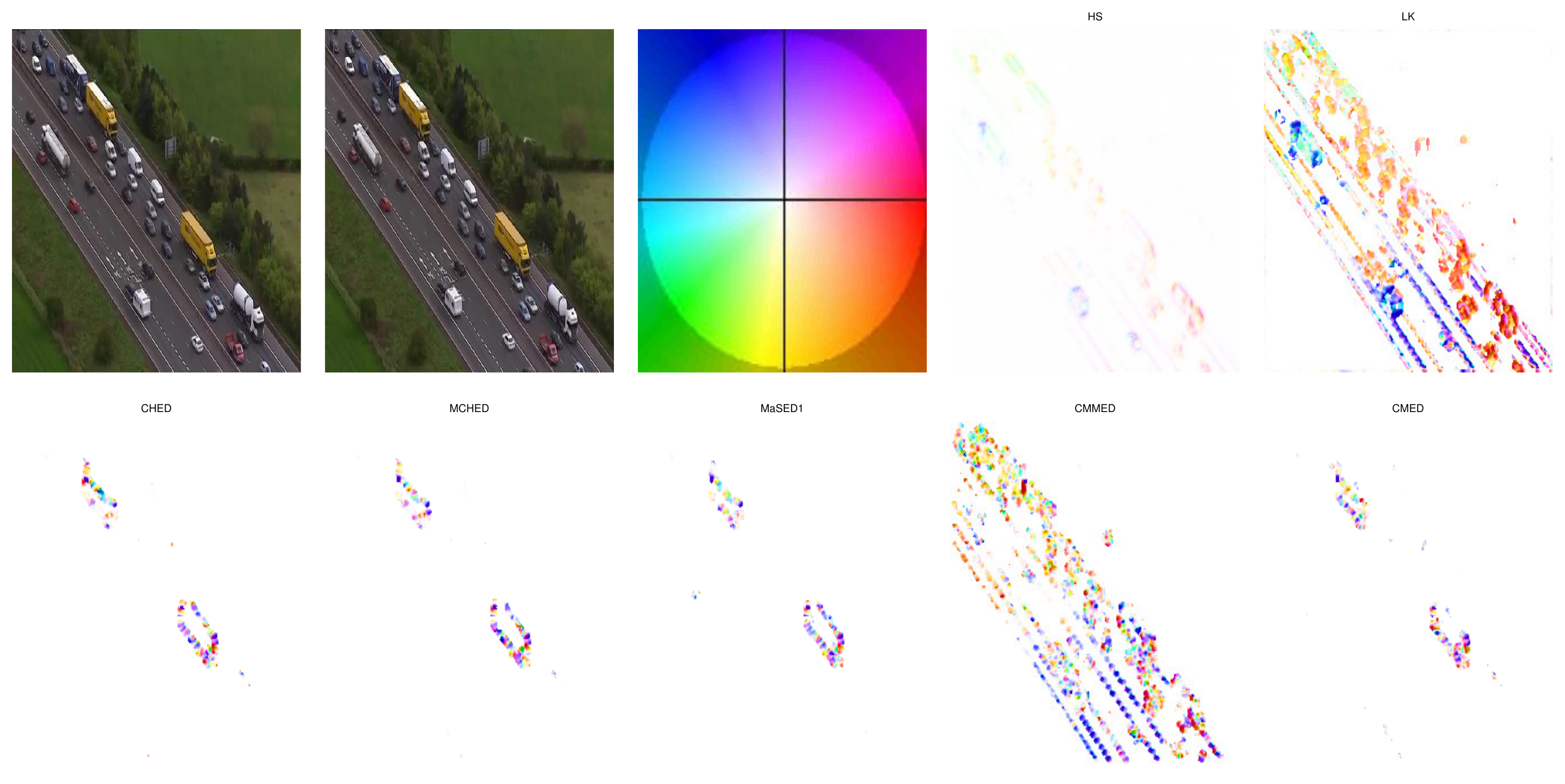}
					\caption{Upper left to right: two frames from the traffic  sequence, flow coding pattern, 	Horn/Schunk result, Lucas/Kanade result. Bottom left to right: color optical flow of
						yellow color detected by CHED, MCHED, MaSED1, CMMED \cite{demarcq2011color}, CMED \cite{demarcq2011color}, respectively.}
					\label{flow}
				\end{figure*}

\begin{figure*}[!ht]
					\centering
					\subfloat[]{\includegraphics[width=2.26in]{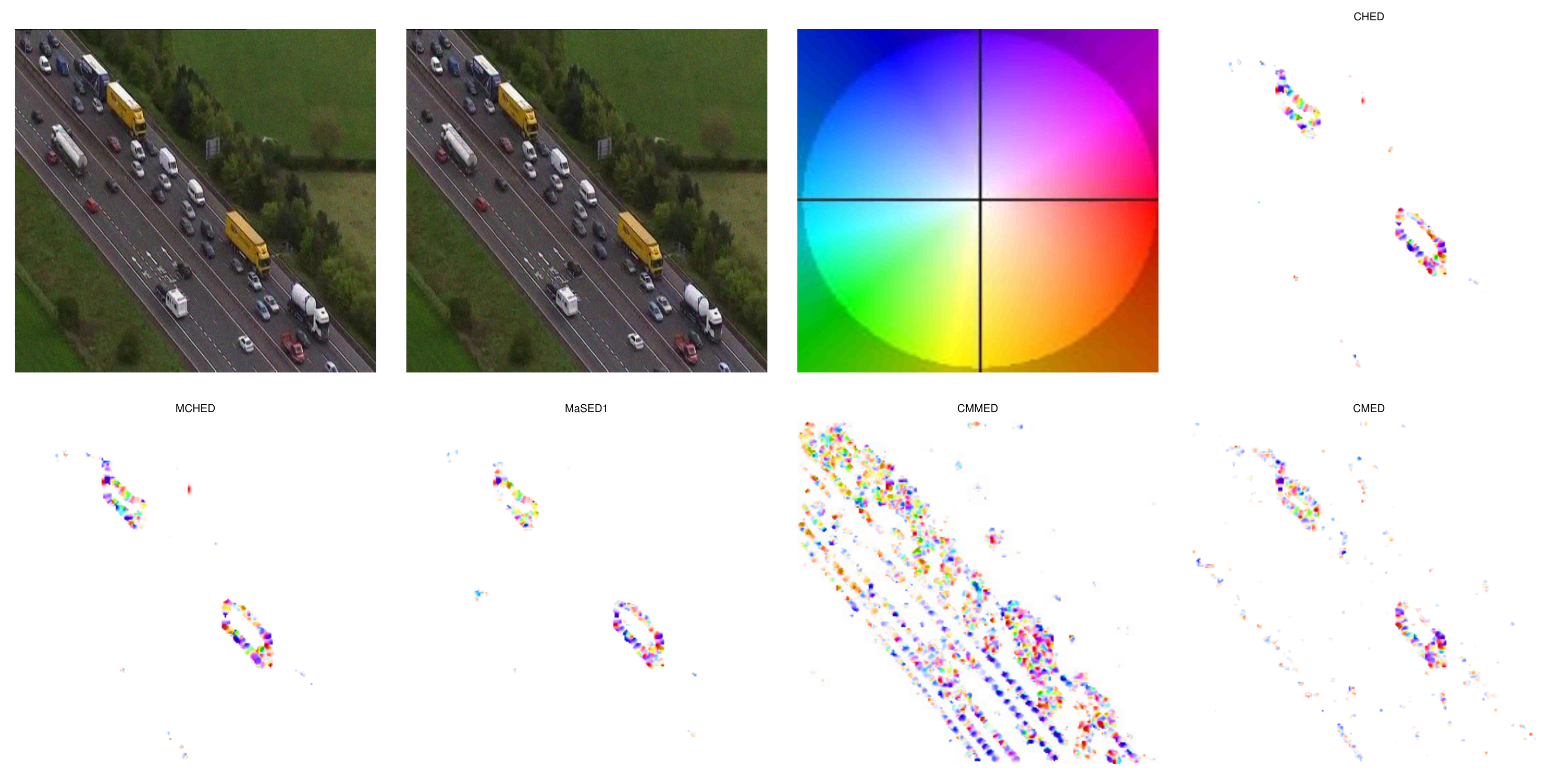}%
					}
					\hfil
					\subfloat[]{\includegraphics[width=2.26in]{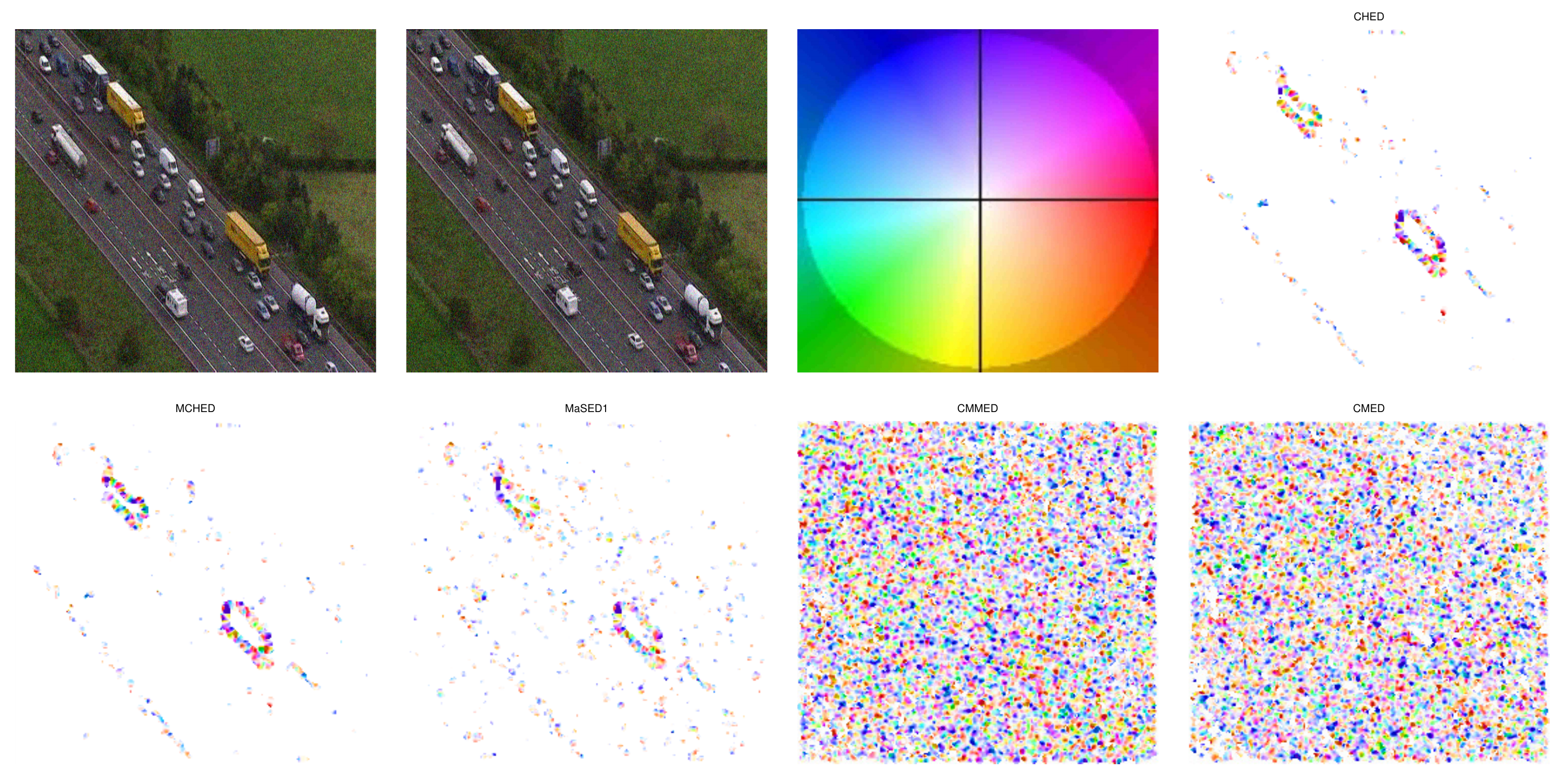}%
					}\\
					\subfloat[]{\includegraphics[width=2.26in]{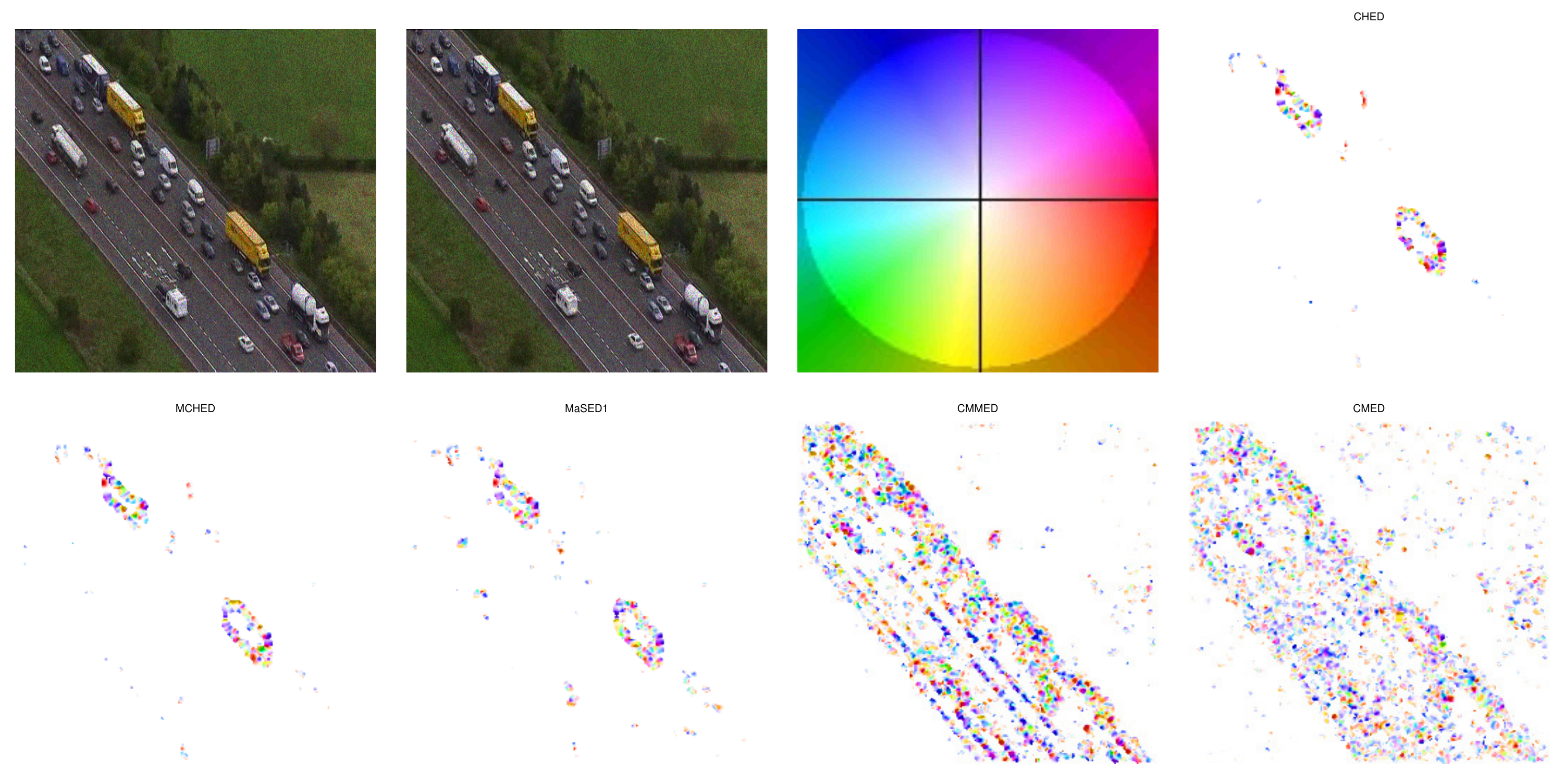}%
					}\hfil
					\subfloat[]{\includegraphics[width=2.26in]{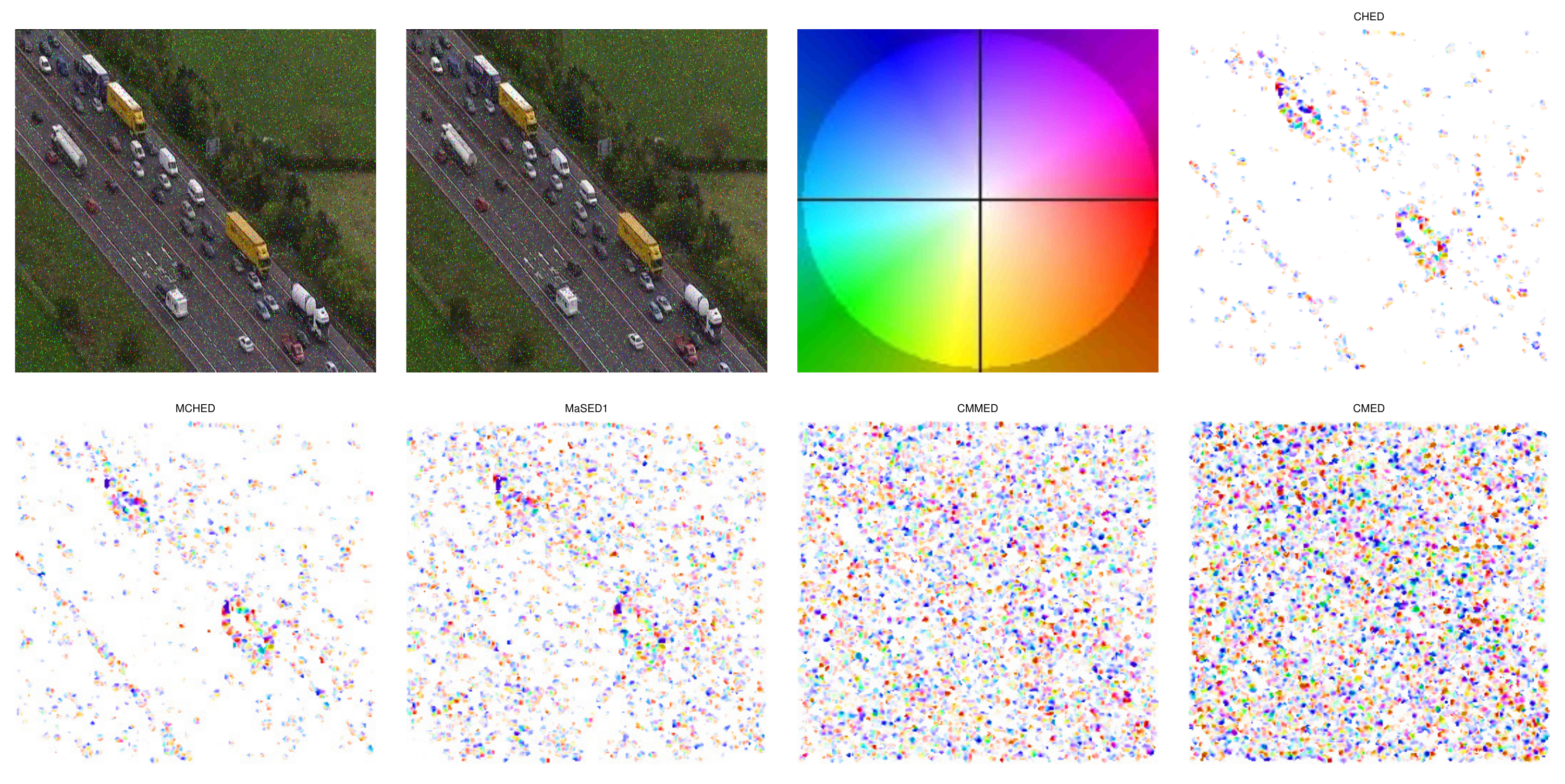}%
					}
					\caption{From left to right, top to bottom:  the results of the two frames from the traffic  sequence corrupted by Possion(a) , Gaussian(b), Speckle(c),  Salt and pepper noises(d) ,respectively. In each sub figure, from left to right, top to bottom: two frames from the traffic sequence corrupted by noises, flow coding pattern,  color optical flow of yellow color detected by CHED, MCHED, MaSED1, CMMED \cite{demarcq2011color}, CMED \cite{demarcq2011color}, respectively.}
					\label{flownoise}
				\end{figure*}

	It is significant to note that the suggested CCHS has an excellent anti-noise performance and is good at detecting the edges for a certain color. In light of this, we will  detect the motion vectors of objects with a specified color in videos. The proposed approach  relies on the local Lucas/Kanade (LK)   method \cite{lucas1981iterative} with a pretreatment. Before  detecting the motion vectors, we perform the edge detection operators on     the frames of the video. Consequently, a pretreated video whose frames are  edge maps   is obtained. We then perform the LK method on this preprocessed video to detect the motion vectors of objects. A low-resolution video of traffic\footnote{\url{https://www.youtube.com/watch?v=goVjVVaLe10&ab_channel=Silverfox1100}} is used as an example to show the validity of the proposed method.  As can be seen from    Figure \ref{flow}  that
	the results	produced by the Horn/Schunk method    and   the local LK method   are not very satisfactory. They can not figure out the vector field of the yellow trucks. Figures \ref{flow}, \ref{flownoise} show that the color
	optical flow obtained by the proposed method  appears to be   accurate. Moreover, it  is robust against   Possion, Gaussian, and speckle noises. 
	
	\section{Conclusions}\label{hxx5}
	The CCHS improves the original image's color information in addition to its vertical and horizontal structural information. The color edge feature can be retrieved by the CCHS in a comprehensive manner by making use of the potent hyper-complex modeling of color. The relationships between the components of the CCHS in high dimensional spaces are first obtained in this study using the generalized Cauchy-Riemann equations. These relationships are helpful when developing edge detection techniques from various angles.
	Second, we provide a variety of edge detection techniques for a specific color, all based on the regional characteristics of CCHS. Last but not least, comparative tests on synthetic and real-world images demonstrate the superiority of the suggested methods.
	
	\section{Acknowledgements}
	Xiaoxiao Hu was supported by the Research Development Foundation of Wenzhou Medical University (QTJ18012). This work was partially supported by  University of Macau (File no. MYRG2019-00039-FST).

	\appendices
	
	\section{The proof of Theorem \ref{crlem}}
	\label{a_sec1}
	\begin{proof}
		For $i=1,2,3$, let
		\begin{align*} 
			A_{4}^{i}(z_{1},z_{2})  &=   [f_{i}*K^{QP}_{y_1,y_2}](x_1,x_2),  \\
			A_{5}^{i}(z_{1},z_{2})& =   [f_{i}*K^{PQ}_{y_1,y_2}](x_1,x_2),\\
			A_{6}^{i}(z_{1},z_{2})& =   [f_{i}*K^{QP}_{y_1,y_2}](x_1,x_2).
		\end{align*}
		Define the  quaternion-valued functions as follows:
		\begin{align*} 
			f_{iq}=& A_{i}+\qi A_{4}^{i}+A_{5}^{i}\qj 
			+ \k A_{6}^{i},\quad i=1,2,3.
		\end{align*}
		It is easy to verify that $f_{iq}(z_1,z_2)$, ($i=1,2,3$) are quaternion Hardy functions. It follows that
		\begin{equation*} 
			\frac{\partial}{\partial \overline{z_{1}}}f_{iq}( z_1, z_2)=0,\quad
			f_{iq}( z_1, z_2)\frac{\partial}{\partial \overline{z_{2}}}=0.
		\end{equation*}
		Therefore
		\begin{align*} 
			\frac{\partial(A_i)}{\partial x_1}&= \frac{\partial(A_{4}^{i})}{\partial y_1},\qquad
			\frac{\partial(A_i)}{\partial x_2}=\frac{\partial(A_{5}^{i})}{\partial y_2},\\
			\frac{\partial(A_{4}^{i})}{\partial x_1}&= -\frac{\partial(A_i)}{\partial y_1},\qquad
			\frac{\partial(A_{4}^{i})}{\partial x_2}=\frac{\partial(A_{6}^{i})}{\partial y_2},\\
			\frac{\partial(A_{5}^{i})}{\partial x_1}&= \frac{\partial(A_{6}^{i})}{\partial y_1},\qquad
			\frac{\partial(A_{5}^{i})}{\partial x_2}=-\frac{\partial(A_i)}{\partial y_2},\\
			\frac{\partial(A_{6}^{i})}{\partial x_1}&= -\frac{\partial(A_{5}^{i})}{\partial y_1},\qquad
			\frac{\partial(A_{6}^{i})}{\partial x_2}=-\frac{\partial(A_{4}^{i})}{\partial y_2}, \quad i=1,2,3.
		\end{align*}
		Note that
		$A_{4}  = \sum_{i=1}^{3}A_{4}^{i},
		A_{5} =\sum_{i=1}^{3} A_{5}^{i},
		A_{6}= \sum_{i=1}^{3}A_{6}^{i}.$
		which completes the proof.
	\end{proof}

	\section{The derivation of Method \ref{m2}, MCHED}
	
	\label{a_sec2}
	Using  Theorem \ref{crlem}, we can substitute the derivative of $ {\rm{Sc}}[f_{cq}\nu]$ with respect to $x_1$ and $x_2$  by the scale derivatives with regard to   $y_1$ and $y_2 $, respectively.
	By simple computation, we can see that
	\begin{align} 
		\frac{\partial {\rm{Sc}}[f_{cq}\nu]}{\partial x_1}&= a\frac{\partial(A_{4}^{1})}{\partial y_1}+b\frac{\partial(A_{4}^{2})}{\partial y_1} +c\frac{\partial(A_{4}^{3})}{\partial y_1}, \label{mmed1} \\
		\frac{\partial {\rm{Sc}}[f_{cq}\nu]}{\partial x_2}&= a\frac{\partial(A_{5}^{1})}{\partial y_2}+b\frac{\partial(A_{5}^{2})}{\partial y_2} +c\frac{\partial(A_{5}^{3})}{\partial y_2}, \label{mmed2}
	\end{align}
	\begin{align} 
		\frac{\partial  |{\rm{Bi}}[f_{cq}\nu]|}{\partial x_1}
		=&
		\frac{1}{|{\rm{Bi}}[f_{cq}\nu]|}
		\bigg((aA_2-bA_1)\big(a\frac{\partial(A_{4}^{2})}{\partial y_1}-b\frac{\partial(A_{4}^{1})}{\partial y_1} \big)\nonumber\\
		+&(aA_3-cA_1)\big(a\frac{\partial(A_{4}^{3})}{\partial y_1}-c\frac{\partial(A_{4}^{1})}{\partial y_1} \big)\nonumber\\
		+&(bA_3-cA_2)\big(b\frac{\partial(A_{4}^{3})}{\partial y_1}-c\frac{\partial(A_{4}^{2})}{\partial y_1} \big)\nonumber\\
		+&(a^2+b^2+c^2)\big( -A_4 \frac{\partial(A_{1}+A_{2}+A_{3})}{\partial y_1} \nonumber\\
		-&bA_5 \frac{\partial(A_{6})}{\partial y_1} 
		-A_6 \frac{\partial(A_{5})}{\partial y_1} \big)\bigg)^{1/2},  \label{mmed3}
	\end{align}
	and
	\begin{align} 
		\frac{\partial  |{\rm{Bi}}[f_{cq}\nu]|}{\partial x_2}
		=&
		\frac{1}{|{\rm{Bi}}[f_{cq}\nu]|}
		\bigg((aA_2-bA_1)\big(a\frac{\partial(A_{5}^{2})}{\partial y_2}-b\frac{\partial(A_{5}^{1})}{\partial y_2} \big)\nonumber\\
		+&(aA_3-cA_1)\big(a\frac{\partial(A_{5}^{3})}{\partial y_2}-c\frac{\partial(A_{5}^{1})}{\partial y_2} \big)\nonumber\\
		+&(bA_3-cA_2)\big(b\frac{\partial(A_{5}^{3})}{\partial y_2}-c\frac{\partial(A_{5}^{2})}{\partial y_2} \big)\nonumber\\
		+&(a^2+b^2+c^2)\big(A_4  \frac{\partial(A_{6})}{\partial y_2} -A_6 \frac{\partial(A_{4})}{\partial y_2}\nonumber\\
		-& bA_5 \frac{\partial(A_{1}+A_{2}+A_{3})}{\partial y_2}  
		\big)\bigg)^{1/2}.  \label{mmed4}
	\end{align}

	Let  $ B_1$,  $B_2 $, $ C_1$ and  $C_2 $
	denote the right hand sides of  Eqs.  (\ref{mmed1}), (\ref{mmed2}), (\ref{mmed3}) and (\ref{mmed4}), respectively. Using   Eqs.  (\ref{mmed1}), (\ref{mmed2}), (\ref{mmed3}) and (\ref{mmed4}),  the  $ \frac{\partial  \theta (x_1,x_2,y_1,y_2)}{\partial x_1}$
	and  $ \frac{\partial  \theta (x_1,x_2,y_1,y_2)}{\partial x_2}$   can be  replaced  by the scale derivatives of the components of the CCHS with respect to $y_1$ and $y_2$.
	That is,
	\begin{align}
		\frac{\partial {\theta}}{\partial x_1}=&
		\frac{1}{M^2}\bigg(C_1{\rm{Sc}}[f_{cq}\nu]-  B_1 |{\rm{Bi}}[f_{cq}\nu]| \bigg), \label{mp11} \\ 
		\frac{\partial  {\theta} }{\partial x_2}=&
		\frac{1}{M^2}\bigg(C_2 {\rm{Sc}}[f_{cq}\nu]-  B_2 |{\rm{Bi}}[f_{cq}\nu]| \bigg).  \label{mp12}
	\end{align}
	Let $D_1$, $D_2$ denote the right hand sides of Eqs.  (\ref{mp11}) and (\ref{mp12}) respectively.
	Then we obtain the   representation of the squared local contrast of $ \mathbf{\mathbf{g}_{f_{cq}\nu}}$  in terms of the  derivatives with respect to the scales  $y_1, y_2$. That is,
	\begin{equation*}
		\widetilde{S}(p,\mathbf{n})=\sum_{i=1}^{2}\sum_{k=1}^{2}\widetilde{\gamma}_{ik}n_in_k,
	\end{equation*}
	where $  \widetilde{\gamma}_{ik}= \left(  B_i ,D_i \right) \cdot  \left(   B_k ,
	D_k \right) $. 
	We develop the MCHED  approach based on this alternate illustration of the squared local contrast of $\mathbf{\mathbf{g}_{f_{cq}\nu}}$.

	\section{The derivation of Method \ref{m5}, MaSED3}
	\label{a_sec4}

	Using   Theorem \ref{crlem}, we can substitute the scale derivatives of $ {\rm{Sc}}[f_{cq}\nu] $ and $ \theta$ with respect to $y_1,y_2$ by the spatial  derivatives with respect to $x_1,x_2 $, respectively. By straightforward calculation,  we have that 
	\begin{align} 
		\frac{\partial {\rm{Sc}}[f_{cq}\nu]}{\partial y_1} =& -a\frac{\partial(A_{4}^{1})}{\partial x_1}-b\frac{\partial(A_{4}^{2})}{\partial x_1} -c\frac{\partial(A_{4}^{3})}{\partial x_1},\label{rp7}\\
		\frac{\partial {\rm{Sc}}[f_{cq}\nu]}{\partial y_2} =&- a\frac{\partial(A_{5}^{1})}{\partial x_2}-b\frac{\partial(A_{5}^{2})}{\partial x_2} -c\frac{\partial(A_{5}^{3})}{\partial x_2},\label{rp8}
	\end{align}
	\begin{align} 
		\frac{\partial  |{\rm{Bi}}[f_{cq}\nu]|}{\partial y_1} =&
		\frac{1}{|{\rm{Bi}}[f_{cq}\nu]|}
		\bigg((aA_2-bA_1)
		\big( b\frac{\partial(A_{4}^{1})}{\partial x_1}-a\frac{\partial(A_{4}^{2})}{\partial x_1} \big)\nonumber\\
		+&(aA_3-cA_1)\big(-a\frac{\partial(A_{4}^{3})}{\partial x_1}+c\frac{\partial(A_{4}^{1})}{\partial x_1} \big)\nonumber\\
		+&(bA_3-cA_2)\big(-b\frac{\partial(A_{4}^{3})}{\partial x_1}+c\frac{\partial(A_{4}^{2})}{\partial x_1} \big)\nonumber\\
		+&(a^2+b^2+c^2)\big(A_4  \frac{\partial(A_{1}+A_{2}+A_{3})}{\partial x_1} \nonumber\\
		-&bA_5 \frac{\partial(A_{6})}{\partial x_1} 
		+A_6 \frac{\partial(A_{5})}{\partial x_1} \big)\bigg)^{1/2},\label{rp9}
	\end{align}
	and
	\begin{align} 
		\frac{\partial  |{\rm{Bi}}[f_{cq}\nu]|}{\partial y_2}=&	
		\frac{1}{|{\rm{Bi}}[f_{cq}\nu]|} \bigg((aA_2-bA_1)\big( b\frac{\partial(A_{5}^{1})}{\partial x_2}-a\frac{\partial(A_{5}^{2})}{\partial x_2} \big)\nonumber\\
		+&(aA_3-cA_1)\big(c\frac{\partial(A_{5}^{1})}{\partial x_2}-a\frac{\partial(A_{5}^{3})}{\partial x_2}  \big)\nonumber\\
		+&(bA_3-cA_2)\big(c\frac{\partial(A_{5}^{2})}{\partial x_2} -b\frac{\partial(A_{5}^{3})}{\partial x_2} \big)\nonumber\\
		+&(a^2+b^2+c^2)\big(A_6 \frac{\partial(A_{4})}{\partial x_2} -A_4  \frac{\partial(A_{6})}{\partial x_2} \nonumber\\
		+&bA_5\frac{\partial(A_{1}+A_2+A_3)}{\partial x_2}\big)\bigg)^{1/2}.\label{rp10} 
	\end{align}
	By the definition of $\theta$, we have that
	\begin{align} 
		\frac{\partial  \theta }{\partial y_1} =&
		\frac{1}{M^2}\bigg( \frac{\partial  |{\rm{Bi}}[f_{cq}\nu]|}{\partial y_1}{\rm{Sc}}[f_{cq}\nu]-  \frac{\partial {\rm{Sc}}[f_{cq}\nu]}{\partial y_1} |{\rm{Bi}}[f_{cq}\nu]| \bigg),   \label{rp11}\\
		\frac{\partial  \theta }{\partial y_2} =&
		\frac{1}{M^2}\bigg( \frac{\partial  |{\rm{Bi}}[f_{cq}\nu]|}{\partial y_2}{\rm{Sc}}[f_{cq}\nu]-  \frac{\partial {\rm{Sc}}[f_{cq}\nu]}{\partial y_2} |{\rm{Bi}}[f_{cq}\nu]| \bigg).  \label{rp12}
	\end{align}
	
	Substituting $\frac{\partial  |{\rm{Bi}}[f_{cq}\nu]|}{\partial y_i}$  and  $\frac{\partial {\rm{Sc}}[f_{cq}\nu]}{\partial y_i}$, ($i=1,2$) in   Eqs. (\ref{rp11}) and (\ref{rp12}) with the right hand sides of 
	Eqs. (\ref{rp7}), (\ref{rp8}), (\ref{rp11}) and  (\ref{rp12}) . Using the relationship between  $\mathbf{\mathbf{g}_{f_{cq}\nu}}$ and $\theta$, we can express  the  scale derivatives $\frac{\partial\mathbf{\mathbf{g}_{f_{cq}\nu}}}{\partial y_1}$ and $\frac{\partial\mathbf{\mathbf{g}_{f_{cq}\nu}}}{\partial y_2}$ in terms of the spatial derivatives.  As a result, we get the    matrix  $I_3(p)$, which is the squared norm represented using only the spatial derivatives.

	\bibliographystyle{IEEEtran}
	\bibliography{IEEEabrv,mybibfile0330}

\begin{thebibliography}{10}
\providecommand{\url}[1]{#1}
\csname url@samestyle\endcsname
\providecommand{\newblock}{\relax}
\providecommand{\bibinfo}[2]{#2}
\providecommand{\BIBentrySTDinterwordspacing}{\spaceskip=0pt\relax}
\providecommand{\BIBentryALTinterwordstretchfactor}{4}
\providecommand{\BIBentryALTinterwordspacing}{\spaceskip=\fontdimen2\font plus
\BIBentryALTinterwordstretchfactor\fontdimen3\font minus
  \fontdimen4\font\relax}
\providecommand{\BIBforeignlanguage}[2]{{%
\expandafter\ifx\csname l@#1\endcsname\relax
\typeout{** WARNING: IEEEtran.bst: No hyphenation pattern has been}%
\typeout{** loaded for the language `#1'. Using the pattern for}%
\typeout{** the default language instead.}%
\else
\language=\csname l@#1\endcsname
\fi
#2}}
\providecommand{\BIBdecl}{\relax}
\BIBdecl

\bibitem{xu2022weakly}
Y.~Xu, X.~Yu, J.~Zhang, L.~Zhu, and D.~Wang, ``Weakly supervised {RGB-D}
  salient object detection with prediction consistency training and active
  scribble boosting,'' \emph{IEEE Transactions on Image Processing}, vol.~31,
  pp. 2148--2161, 2022.

\bibitem{arthur2007k}
D.~Arthur and S.~Vassilvitskii, ``K-means++ the advantages of careful
  seeding,'' in \emph{Proceedings of the eighteenth annual ACM-SIAM symposium
  on Discrete algorithms}, 2007, pp. 1027--1035.

\bibitem{laaksonen1996classification}
J.~Laaksonen and E.~Oja, ``Classification with learning k-nearest neighbors,''
  in \emph{Proceedings of international conference on neural networks
  (ICNN'96)}, vol.~3.\hskip 1em plus 0.5em minus 0.4em\relax IEEE, 1996, pp.
  1480--1483.

\bibitem{shi2007quaternion}
L.~Shi and B.~Funt, ``Quaternion color texture segmentation,'' \emph{Computer
  Vision and image understanding}, vol. 107, no. 1-2, pp. 88--96, 2007.

\bibitem{sathya2021color}
P.~Sathya, R.~Kalyani, and V.~Sakthivel, ``Color image segmentation using
  {Kapur}, {Otsu} and minimum cross entropy functions based on exchange market
  algorithm,'' \emph{Expert Systems with Applications}, vol. 172, p. 114636,
  2021.

\bibitem{burke2021edge}
J.~Burke and S.~King, ``Edge tracing using {Gaussian} process regression,''
  \emph{IEEE Transactions on Image Processing}, vol.~31, pp. 138--148, 2022.

\bibitem{feng2021deep}
Y.~Feng, A.~Hafiane, and H.~Laurent, ``A deep learning based multiscale
  approach to segment the areas of interest in whole slide images,''
  \emph{Computerized Medical Imaging and Graphics}, vol.~90, p. 101923, 2021.

\bibitem{meng2023multiscale}
K.~Meng, X.~Dong, H.~Shan, and S.~Xia, ``Multiscale hierarchical attention
  fusion network for edge detection,'' \emph{International Journal of Ad Hoc
  and Ubiquitous Computing}, vol.~42, no.~1, pp. 1--11, 2023.

\bibitem{chen2023edge}
C.~Chen, C.~Wang, B.~Liu, C.~He, L.~Cong, and S.~Wan, ``Edge intelligence
  empowered vehicle detection and image segmentation for autonomous vehicles,''
  \emph{IEEE Transactions on Intelligent Transportation Systems}, 2023.

\bibitem{bulow1999novel}
T.~B{\"u}low and G.~Sommer, ``A novel approach to the {2D} analytic signal,''
  in \emph{Computer Analysis of Images and Patterns}, F.~Solina and
  A.~Leonardis, Eds.\hskip 1em plus 0.5em minus 0.4em\relax Berlin, Heidelberg:
  Springer, 1999, pp. 25--32.

\bibitem{bernstein2013generalized}
S.~Bernstein, J.-L. Bouchot, M.~Reinhardt, and B.~Heise, \emph{Generalized
  Analytic Signals in Image Processing: Comparison, Theory and
  Applications}.\hskip 1em plus 0.5em minus 0.4em\relax Basel: Springer Basel,
  2013, pp. 221--246.

\bibitem{pei2008short}
S.-C. Pei, J.-J. Ding, J.-D. Huang, and G.-C. Guo, ``Short response {Hilbert}
  transform for edge detection,'' in \emph{APCCAS 2008-2008 IEEE Asia Pacific
  Conference on Circuits and Systems}.\hskip 1em plus 0.5em minus 0.4em\relax
  IEEE, 2008, pp. 340--343.

\bibitem{kou2017envelope}
K.~I. Kou, M.-S. Liu, J.~P. Morais, and C.~Zou, ``Envelope detection using
  generalized analytic signal in {2D} {QLCT} domains,'' \emph{Multidimensional
  Systems and Signal Processing}, vol.~28, pp. 1343--1366, 2017.

\bibitem{kou2020plancherel}
K.~I. Kou, M.-S. Liu, and C.~Zou, ``Plancherel theorems of quaternion hilbert
  transforms associated with linear canonical transforms,'' \emph{Advances in
  Applied Clifford Algebras}, vol.~30, pp. 1--23, 2020.

\bibitem{felsberg2001monogenic}
M.~Felsberg and G.~Sommer, ``The monogenic signal,'' \emph{IEEE transactions on
  signal processing}, vol.~49, no.~12, pp. 3136--3144, 2001.

\bibitem{felsberg2004monogenic}
------, ``The monogenic scale-space: A unifying approach to phase-based image
  processing in scale-space,'' \emph{Journal of Mathematical Imaging and
  vision}, vol.~21, pp. 5--26, 2004.

\bibitem{yang2018edge}
Y.~Yang, K.~I. Kou, and C.~Zou, ``Edge detection methods based on modified
  differential phase congruency of monogenic signal,'' \emph{Multidimensional
  Systems and Signal Processing}, vol.~29, pp. 339--359, 2018.

\bibitem{hu2018phase}
X.-X. Hu and K.~I. Kou, ``Phase-based edge detection algorithms,''
  \emph{Mathematical Methods in the Applied Sciences}, vol.~41, no.~11, pp.
  4148--4169, 2018.

\bibitem{demarcq2011color}
G.~Demarcq, L.~Mascarilla, M.~Berthier, and P.~Courtellemont, ``The color
  monogenic signal: Application to color edge detection and color optical
  flow,'' \emph{Journal of Mathematical Imaging and Vision}, vol.~40, pp.
  269--284, 2011.

\bibitem{brackx1982clifford}
F.~Brackx, R.~Delanghe, and F.~Sommen, \emph{Clifford Analysis}, ser. Research
  Notes in Mathematics.\hskip 1em plus 0.5em minus 0.4em\relax Boston, London,
  Melbourne: Pitman Advanced Publishing Company, 1982, vol.~76.

\bibitem{jiang2020controllability}
B.~X. Jiang, Y.~Liu, K.~I. Kou, and Z.~Wang, ``Controllability and
  observability of linear quaternion-valued systems,'' \emph{Acta Mathematica
  Sinica, English Series}, vol.~36, no.~11, pp. 1299--1314, 2020.

\bibitem{xia2020penalty}
Z.~Xia, Y.~Liu, J.~Lu, J.~Cao, and L.~Rutkowski, ``Penalty method for
  constrained distributed quaternion-variable optimization,'' \emph{IEEE
  Transactions on Cybernetics}, vol.~51, no.~11, pp. 5631--5636, 2020.

\bibitem{hu2017quaternion}
X.-X. Hu and K.~I. Kou, ``Quaternion {Fourier} and linear canonical inversion
  theorems,'' \emph{Mathematical Methods in the Applied Sciences}, vol.~40,
  no.~7, pp. 2421--2440, 2017.

\bibitem{lovelock1989tensors}
D.~Lovelock and H.~Rund, \emph{Tensors, differential forms, and variational
  principles}.\hskip 1em plus 0.5em minus 0.4em\relax New York: Dover, 1989.

\bibitem{cumani1991edge}
A.~Cumani, ``Edge detection in multispectral images,'' \emph{CVGIP: Graphical
  models and image processing}, vol.~53, no.~1, pp. 40--51, 1991.

\bibitem{evans2006morphological}
A.~N. Evans and X.~U. Liu, ``A morphological gradient approach to color edge
  detection,'' \emph{IEEE Transactions on Image Processing}, vol.~15, no.~6,
  pp. 1454--1463, 2006.

\bibitem{kreyszig1991differential}
E.~Kreyszig, \emph{Differential Geometry}.\hskip 1em plus 0.5em minus
  0.4em\relax New York: Dover, 1991.

\bibitem{tschumperle2006fast}
D.~Tschumperl{\'e}, ``Fast anisotropic smoothing of multi-valued images using
  curvature-preserving {PDE's},'' \emph{International Journal of Computer
  Vision}, vol.~68, pp. 65--82, 2006.

\bibitem{wang2004image}
Z.~Wang, A.~C. Bovik, H.~R. Sheikh, and E.~P. Simoncelli, ``Image quality
  assessment: from error visibility to structural similarity,'' \emph{IEEE
  transactions on image processing}, vol.~13, no.~4, pp. 600--612, 2004.

\bibitem{zhang2011fsim}
L.~Zhang, L.~Zhang, X.~Mou, and D.~Zhang, ``{FSIM}: A feature similarity index
  for image quality assessment,'' \emph{IEEE transactions on Image Processing},
  vol.~20, no.~8, pp. 2378--2386, 2011.

\bibitem{abdou1979quantitative}
I.~E. Abdou and W.~K. Pratt, ``Quantitative design and evaluation of
  enhancement/thresholding edge detectors,'' \emph{Proceedings of the IEEE},
  vol.~67, no.~5, pp. 753--763, 1979.

\bibitem{plataniotis2000color}
K.~Plataniotis and A.~N. Venetsanopoulos, \emph{Color image processing and
  applications}.\hskip 1em plus 0.5em minus 0.4em\relax New York:
  Springer-Verlag, 2000.

\bibitem{androutsos1998color}
P.~Androutsos, D.~Androutsos, K.~N. Plataniotis, and A.~N. Venetsanopoulos,
  ``Color edge detectors: a subjective analysis,'' in \emph{Nonlinear Image
  Processing IX}, vol. 3304.\hskip 1em plus 0.5em minus 0.4em\relax SPIE, 1998,
  pp. 260--267.

\bibitem{martin2001database}
D.~Martin, C.~Fowlkes, D.~Tal, and J.~Malik, ``A database of human segmented
  natural images and its application to evaluating segmentation algorithms and
  measuring ecological statistics,'' in \emph{Proceedings Eighth IEEE
  International Conference on Computer Vision. ICCV 2001}, vol.~2.\hskip 1em
  plus 0.5em minus 0.4em\relax IEEE, 2001, pp. 416--423.

\bibitem{lucas1981iterative}
B.~D. Lucas and T.~Kanade, ``An iterative image registration technique with an
  application to stereo vision,'' in \emph{IJCAI'81: 7th international joint
  conference on Artificial intelligence}, vol.~2, 1981, pp. 674--679.

\end{thebibliography}

	\vfill

	\begin{table*}[!t] \scriptsize
		\caption{The quantitative indexes of the edge operators  on the synthetic image.}\label{tab2}
		\centering
		\begin{tabular}{|c|c|c|c|c|c|c|c|c|}
			\hline
			\multicolumn{2}{|c|}{Method} & CHED &MCHED & MaSED1 & MaSED2 & MaSED3 & CMED \cite{demarcq2011color} & CMMED \cite{demarcq2011color} \\ \hline
			& Blue edge  & 0.9944 & \textbf{0.9945} & \textbf{0.9945} & 0.9789 &  0.9937 & 0.9941 & 0.9942 \\
			SSIM &Red edge &0.9572 & \textbf{0.9582} & 0.9572 & 0.9265 & 0.9263 & 0.9440 & 0.9356 \\
			& Yellow edge & \textbf{0.9742} & \textbf{0.9742} &  0.9741 & 0.9654 & 0.9656 & 0.9562 & 0.9740   \\ \hline 
			\hline
			& Blue edge & \textbf{0.9994}& \textbf{0.9994} & \textbf{0.9994} & 0.9971& 0.9993 &0.9993 &\textbf{0.9994} \\
			FSIM & Red edge & 0.9932 & 0.9934 & 0.9932 & 0.9878 & 0.9888 & \textbf{0.9959} & 0.9902 \\
			& Yellow edge & 0.9970  & 0.9970 & \textbf{0.9971} & 0.9960 & 0.9961& 0.9832 &0.9962  \\
			\hline\hline
			& Blue edge & 0.9763 & \textbf{0.9800} & 0.9763 & 0.9448 & 0.9602 & 0.9643 & 0.9613 \\ 
			F & Red edge & \textbf{0.9263} & 0.9254 & \textbf{0.9263} & 0.8855 & 0.8868 &  0.8886 & 0.8980 \\
			& Yellow edge & 0.9243 & \textbf{0.9269} & 0.9206  & 0.8928 & 0.9021 & 0.9254 & 0.9141 \\  \hline \hline
			& Blue edge & \textbf{30.1714} & 30.1130 & \textbf{30.1714} & 24.5287 & 30.0357 & 30.1030 & 30.1163 \\
			PSNR &  Red edge & 21.4623& \textbf{21.5847} &21.4623 &19.0961& 19.0799& 21.3328 &19.7606 \\
			& Yellow edge & 23.7088& 23.6623 &23.7088& 22.4129& 22.4477& 21.5847 & \textbf{23.8859} \\   \hline 
		\end{tabular}
	\end{table*}

	\begin{table*}[!t]\scriptsize
		\caption{The quantitative indexes of the edge operators  on the  real-world images  corrupted by Poisson noise.}\label{tab4}
		\centering
		\begin{tabular}{|c|c|c|c|c|c|c|c|c|c|c|c|c|}
			\hline
			& \multicolumn{4}{c|}{Flowers: SNR=20.778}&  \multicolumn{4}{c|}{Fence: SNR=21.9249}&\multicolumn{4}{c|}{Blue parts of Woman: SNR=21.3450}\\
			\hline
			& SSIM  &FSIM & F & PSNR  & SSIM & FSIM & F & PSNR  & SSIM & FSIM & F & PSNR \\
			CHED & 0.8733 & 0.9872& 0.9590& 17.4093 & \textbf{0.9775} & 0.9983&0.9798&\textbf{24.7802}&0.9715&0.9800&0.8021&24.6237\\
			MCHED & \textbf{0.8768}& 0.9876& \textbf{0.9642}&17.6225 &0.9757  & \textbf{0.9987}&\textbf{0.9898}&24.2554&0.9765&0.9912&0.9202&25.5403\\
			MaSED1 &0.8674 &\textbf{0.9884} &0.9539 &16.9492  &0.9701 & 0.9982&0.9757&22.9534&\textbf{0.9825}&\textbf{0.9988}&\textbf{0.9719}&\textbf{26.0164}\\
			MaSED2 &0.7820 & 0.9743 &0.8896 &15.8196  &0.9018  & 0.9743 &0.9118&18.7200&0.9639&0.9955&0.8514&23.5860\\
			MaSED3&0.8583& \textbf{0.9844}& 0.9383&\textbf{17.7868}& 0.9558&0.9975& 0.9765&22.0264&0.9669&0.9964&0.8754&23.7715\\
			CMED \cite{demarcq2011color} & 0.7785  &0.9731&0.8711&14.3123&0.9614 &0.9976&0.9707&21.9117&0.9699&0.9831&0.7054&25.0051\\
			CMMED \cite{demarcq2011color} & 0.7196 &0.9650&0.8500&14.1768&0.9472 & 0.9943&0.9596&21.0467&0.9572&0.9676&0.6618&23.8512\\
			\hline\hline
			&\multicolumn{4}{c|}{Red Parts of Woman: SNR=29.3947}&  \multicolumn{4}{c|}{Fireman: SNR=20.7992 }&\multicolumn{4}{c|}{Fish: SNR=19.3275}\\
			\hline
			& SSIM & FSIM& F & PSNR   & SSIM & FSIM & F & PSNR & SSIM & FSIM & F & PSNR \\
			CHED & \textbf{0.9937}& \textbf{0.9996}& \textbf{0.9903}& \textbf{31.0048} & \textbf{0.9829} & \textbf{0.9992}& 0.9831 &25.9965&\textbf{0.9975}&\textbf{0.9998}&\textbf{0.9960}&\textbf{35.1445}\\
			MCHED & 0.9920& \textbf{0.9996}& 0.9886&29.7763  & 0.9800 &0.9989&0.9800&25.4232&0.9963&0.9993&0.9866&33.3936\\
			MaSED1  &0.9916&0.9995 &0.9825 &29.4142& 0.9832 & 0.9991&\textbf{0.9861}&\textbf{26.0965}&0.9962&\textbf{0.9998}&0.9916&32.8500\\
			MaSED2 &0.9752& 0.9978 &0.9629 &24.5855  & 0.9633 & 0.9953&0.9317&22.8756&0.9869&0.9991&0.9791&27.9117\\
			MaSED3& 0.9868&0.9981& 0.9455&28.0364& 0.9703&  0.9977&0.9462&24.1681&0.9908&0.9993&0.9731&29.6522\\
			CMED \cite{demarcq2011color} &0.9850&0.9991&0.9668&26.8294& 0.9659 & 0.9965&0.9326&23.6469&0.9917&0.9967&0.9586&31.2628\\
			CMMED  \cite{demarcq2011color} &0.9456&0.9907&0.9186&24.9426& 0.9800 &0.9956&0.9287&22.3442&0.9810&0.9980&0.9506&26.6726\\
			\hline
		\end{tabular}
	\end{table*}

				\begin{table*}[!t]\scriptsize
					\caption{The quantitative indexes of the edge operators  on the fence image  corrupted by different types of noises. }\label{other}
					\centering
					\begin{tabular}{|c|c|c|c|c|c|c|c|c|c|}
						\hline
						Poisson   & SSIM   &  FSIM   &F  &  PSNR    &   Gaussian  & SSIM   & FSIM  &F  & PSNR  \\ \hline  
						CHED &\textbf{0.9779}& \textbf{0.9989}& 0.9841  &\textbf{24.9633} &CHED &\textbf{0.9637}& \textbf{0.9975}& \textbf{ 0.9582}&\textbf{22.4477}\\
						MCHED    & 0.9764 &  0.9988 & \textbf{0.9848} &  24.4173 & MCHED  & 0.9528 &   0.9887 & 0.9491 & 21.4438 \\
						MaSED1       & 0.9741    & 0.9986   & 0.9769      &23.6469  &MaSED1       & 0.8713&0.9064   &0.8437  &18.5411    \\
						MaSED2     & 0.9299    & 0.9857 & 0.9108       &19.7387  &MaSED2    & 0.6818&0.8941 &0.2019  &14.5912    \\
						MaSED3     & 0.9548    & 0.9972  & 0.9691     &21.9634  & MaSED3   & 0.9034&0.9115&0.8915   &18.6079    \\
						CMED \cite{demarcq2011color}    & 0.7390    & 0.8974    & 0.9662     &16.6389  & CMED \cite{demarcq2011color}  & 0.0936&0.7988 &0.3124 &6.6966    \\
						CMMED \cite{demarcq2011color} & 0.3422   & 0.8732    & 0.9727     &19.8705  &CMMED \cite{demarcq2011color} & 0.0320&0.7931 &0.8823  &6.1165    \\
						NNM  \cite{laaksonen1996classification}  & 0.9433    & 0.9967 & 0.1101      &20.5305  &NNM \cite{laaksonen1996classification} & 0.6680&0.8811 &0.4785  &14.7032    \\
						KMCM \cite{arthur2007k}  & 0.8847    & 0.9808   & 0.8587     &17.0120  & KMCM \cite{arthur2007k} & 0.7411&0.8921 &0.6597 &15.2312    \\
						\hline\hline
						Salt and pepper   & SSIM & FSIM &  F &PSNR &    Speckle  & SSIM & FSIM & F& PSNR\\
						\hline
						CHED &\textbf{0.9573}& \textbf{0.9974}&  \textbf{0.9638}&\textbf{21.6617} &CHED &\textbf{0.9637}& \textbf{ 0.9975}& \textbf{0.9765}& \textbf{ 22.4477}\\
						MCHED    & 0.9470 &  0.9967 &  0.8999 &  20.6907  &MCHED  & 0.7012 &   0.8858 & 0.9710 &  15.2113 \\
						MaSED1       & 0.8386   & 0.9015  & 0.4875     &17.9627 & MaSED1   & 0.1784  &0.8442 & 0.7558 &9.5393   \\
						MaSED2     & 0.6343   & 0.8874   & 0.2736  &13.8818 & MaSED2  & 0.1202&0.8341 & 0.2015 &9.5120    \\
						MaSED3     & 0.8431    & 0.9311  & 0.6512    &17.4023 & MaSED3 & 0.2749&0.8629 & 0.8554  &11.3119    \\
						CMED \cite{demarcq2011color} & 0.0298   & 0.7947    & 0.1968   &6.7191  & CMED \cite{demarcq2011color} & 0.1009&0.8008 & 0.2278 &6.8278   \\
						CMMED \cite{demarcq2011color}   & 0.0171  & 0.7929  & 0.0804     &6.0499  &CMMED \cite{demarcq2011color}  & 0.0654&0.7979 & 0.7283&6.3601    \\
						NNM \cite{laaksonen1996classification}   & 0.5217   & 0.8812  & 0.1470   &14.7505 & NNM \cite{laaksonen1996classification}& 0.6334&0.8812  & 0.0979 &12.8093   \\
						KMCM  \cite{arthur2007k} & 0.4922   & 0.8827  & 0.6008  &13.3777  & KMCM \cite{arthur2007k}  & 0.6839&0.8808 & 0.4379&13.7244    \\
						\hline
					\end{tabular}
				\end{table*}

				
				\begin{table*}[!t] \scriptsize
					\caption{The quantitative indexes of the edge operators  on the traffic flow  corrupted by different types of noises.}\label{tab_flow}
					\centering
					\begin{tabular}{|c|c|c|c|c|c|c|c|}
						\hline
						Poisson (SNR=19.1629) & SSIM   &  FSIM    &  PSNR   &   Gaussian (SNR=9.8607)  & SSIM   & FSIM   & PSNR  \\
						\hline
						CHED &0.9624& 0.9776& 27.5402
						&CHED & \textbf{0.8994}&\textbf{0.8575}& \textbf{23.4021}\\
						MCHED    &0.9638& 0.9737&26.7766
						& MCHED  &0.8733&  0.8064& 23.5886\\
						MaSED1      &\textbf{0.9671}& \textbf{0.9787}& \textbf{27.7952} &MaSED1       & 0.7602&0.7038   &22.3363    \\
						CMED \cite{demarcq2011color} & 0.8999    & 0.8602        &24.0218 & CMED \cite{demarcq2011color}  & 0.0354&0.1644  &11.4869    \\
						CMMED \cite{demarcq2011color}  & 0.7326   & 0.8100        &17.5858  &CMMED \cite{demarcq2011color} & 0.0498&0.3101  &10.8640    \\
						\hline\hline
						Salt and pepper  (SNR=8.0194)  & SSIM & FSIM & PSNR &    Speckle (SNR=13.3416)  & SSIM & FSIM & PSNR\\
						\hline
						CHED &\textbf{0.7148}& \textbf{0.6955}& \textbf{21.0420} &CHED &\textbf{0.9527}& \textbf{0.9473}& \textbf{25.8393}\\
						MCHED    &0.5798& 0.6100& 20.0420 &MCHED  &0.9439&  0.9347& 25.5293\\
						MaSED1       & 0.2458  & 0.4094      &16.7263 &
						MaSED1   & 0.9196 &0.9164  &24.7776   \\
						CMED \cite{demarcq2011color} & 0.0493  & 0.2010   &11.2897 & CMED \cite{demarcq2011color}  & 0.4531  & 0.4972       &16.8616    \\
						CMMED \cite{demarcq2011color}  & 0.0800  & 0.3714      &12.6189 &CMMED \cite{demarcq2011color}  & 0.6048&0.7114  &15.4058   \\
						\hline	
					\end{tabular}	
				\end{table*}
				
				\label{sMED}

			\end{document}